\documentclass[12pt]{iopart}

\usepackage{graphicx}
\usepackage{cite}
\usepackage{color}
\usepackage{multirow}
\usepackage[section]{placeins}
\usepackage{amssymb}

\begin{document}

\setlength{\parindent}{0pt}

\title[Atmospheric-pressure nanosecond discharges]{Effects of excitation voltage pulse shape on the characteristics of atmospheric-pressure nanosecond discharges}

\author{Zolt\'an Donk\'o$^{1,2}$, Satoshi Hamaguchi$^2$, Timo Gans$^3$}

\address{$^1$Institute for Solid State Physics and Optics, Wigner Research Centre for Physics, Hungarian Academy of Sciences, 1121 Budapest, Konkoly Thege Mikl\'os str. 29-33, Hungary\\
$^2$Center for Atomic and Molecular Technologies,
Graduate School of Engineering, Osaka University,
2-1 Yamadaoka, Suita, Osaka 565-0871, Japan\\
$^3$York Plasma Institute, Department of Physics, University of York, Heslington, York, United Kingdom}
\ead{donko.zoltan@wigner.mta.hu}

\begin{abstract}
The characteristics of atmospheric-pressure microdischarges excited by nanosecond high-voltage pulses are investigated in helium-nitrogen mixtures, as a function of the parameters of the excitation voltage pulses. In particular, cases of single-pulse excitation, unipolar and bipolar double-pulse excitation are studied, at different pulse durations, voltage amplitudes, and delay times (for the case of double-pulse excitation). Our investigations are carried out with a particle-simulation code that also comprises the treatment of the VUV resonance radiation in the plasma. The simulations allow gaining insight into the plasma dynamics during and after the excitation pulse, the development and the decay of charged particle density profiles and fluxes. We find a strong dependence of the electron density of the plasma (measured at the end of the excitation pulse) on the electrical input energy into the plasma and a weak influence of the shape of the excitation pulse at the same input energy.
\end{abstract}

\submitto{\PSST}

\maketitle

\section{Introduction}

Transient (short-pulse) discharges have a broad range of technological applications including combustion science \cite{combustion1, combustion2}, aerodynamic flow control \cite{flow1,flow2}, switching applications \cite{Bokhan,Bokhan2}, spectroscopy \cite{spec1,spec2}, synthesis of nanomaterials \cite{nano0}, conversion of CH$_4$/CO$_2$ gases \cite{Scapinello,Lotfalipour} and promising innovative biomedical applications through the generation of reactive species at ambient pressure and temperature with their associated effects on tissue and cells \cite{Beebe,bio0}. Depending on the application these discharges are operated across wide ranges of operating conditions (pressure, voltage amplitude, pulse duration) in various gases and gas mixtures with various electrode configurations \cite{Svetlana,Svetlana2,Starikovskiy,Pai2009}.

The pronounced transient nature of these discharges is a challenge equally for experimental as well as theoretical and computational studies. Different operation modes of nanosecond discharges in helium at sub-atmospheric pressures have been explored in \cite{Kikuchi} and advanced spectroscopic diagnostics tools have been applied to characterise these discharges \cite{Uwe,Stancu,PO}.

We recently developed a fully kinetic particle-based simulation code for a first principle description of short-pulse nanosecond discharges at atmospheric pressure mixtures of helium and nitrogen (at N$_2$ concentrations in the order of 1\%) \cite{Dns}. In this code we have also included the photon treatment of the VUV resonance radiation of helium, based on the approach presented in \cite{Fierro}. This has enabled us to study the general fundamental characteristics of these discharges in a fully kinetic picture \cite{Dns}. We previously explored for typical conditions, the spatiotemporal evolution of charged species densities, reaction rates, and the fluxes of "active" species at the surfaces. We also investigated the behaviour of the electron velocity distribution function and the electron energy probability function, and concluded that these deviate significantly from the Maxwell-Boltzmann form, especially in the cathode region of the discharge. These observations demonstrated the advantages (and uniqueness) of particle simulations [23-30] of similar physical systems. We found that the VUV resonance radiation of He atoms is heavily trapped within the high-pressure gaseous environment and photons are absorbed / re-emitted typically several hundred times before leaving the plasma. Nonetheless, the escaping photons were found to contribute significantly to the electron emission process at the electrodes. For most conditions studied in \cite{Dns}, an increase of around a factor of two of the current pulse peak was observed when VUV photons were included in the simulations, in comparison to those cases when their effect was neglected.
Our previous work \cite{Dns} focused on simplified temporally symmetric single pulse systems. In this paper, we investigate the more realistic and consequently more complex scenarios of different excitation voltage pulse shapes and double voltage pulses. In the latter case, we study both unipolar and bipolar pulses with varying voltage amplitudes as well as the effect of a delay time between the pulses. In the case of single-pulse excitation we also investigate the post-pulse dynamics of the plasma more in detail.

Section 2 briefly summarises the discharge model, the processes accounted for, treatment of the radiation transport and the investigated excitation voltage waveforms. Section 3 presents our simulation results. In section 3.1 the results for single-pulse excitation are presented, while the different cases of double-pulse excitation are covered in section 3.2. Section 4 gives a summary and conclusion of the work.

\section{Simulation method}

We investigate the short-pulse discharges by adopting a particle-based technique, the "standard" PIC/MCC (Particle-in-Cell / Monte Carlo collisions) approach \cite{PIC1,PIC2,PIC3}. Our electrostatic code describes a symmetric bounded plasma that is one dimensional in space and three dimensional in velocity space (1d3v). It traces electrons and three ionic species: He$^+$, He$_2^+$, and N$_2^+$. The code includes, additionally, the photon treatment of the resonance radiation in the plasma. Full details of the simulation model have been presented in \cite{Dns}, below only a short summary of its main features is given.

The model and its computational implementation includes a limited set of elementary processes in order to keep the computational cots within reasonable limits. Time is discretised in the simulation and between consecutive collisions the particles move along trajectories that are defined by their equations of motion. Decision about the occurrence of collisions is based on the comparison of a random number $R_{01}$ (having a uniform distribution over the $[0,1)$ interval) with the collision probability 
\begin{equation}
P = 1- \exp(-\nu_{\rm proj, tot} \Delta t) = 1- \exp(- n_{\rm targ}\, \sigma_{\rm proj, tot}\, g \,\Delta t),
\label{eq:pcoll1}
\end{equation}
where $\nu_{\rm proj, tot}$ is the total collision frequency of the given projectile, $n_{\rm targ}$ is the density of the target species, $\sigma_{\rm proj, tot}$ is the total collision cross section, $g$ is the relative velocity, and $\Delta t$ is the time step for the given projectile. For the ions we take into account the thermal motion of the background gas, while for electrons we use the cold gas approximation.

Electrons can collide with ground-state constituents of the buffer gas, which is helium with a small admixture of molecular nitrogen. For the electron -- atom/molecule collisions we use the cross section sets from \cite{he-cs} and \cite{n2-cs}, respectively. (This latter set is largely based on the Siglo cross section set, accessible now at the LxCat website \cite{siglo}). All electron -- atom/molecule collisions are assumed to result in isotropic scattering. Electrons can collide with He atoms elastically, cause ionisation or excitation. In the latter case, 50\% of the excitation to singlet states is assumed to result in the formation of singlet (2$^1$S) metastable atoms and the other 50\% is assumed to populate the 2$^1$P resonant state, either by direct excitation to these levels or by cascade transitions from higher-lying states. For excitation in the triplet system, 50\% is assumed to lead to the formation of triplet metastables (2$^3$S) \cite{Dns}. The formation of the metastable states is part of a very important ionisation pathway, the Penning ionisation process (see later), while the excitation of the 2$^1$P state leads to the emission of the strong VUV (58.4334 nm) resonance radiation of the He atoms. The set of elementary processes for electron - N$_2$ molecule collisions includes elastic scattering, rotational, vibrational and electronic excitation, as well as dissociation and ionisation.

For the different ionic species, He$^+$ (created by electron impact and Penning ionisation), He$_2^+$ (created via ion conversion), and N$_2^+$ (created via electron impact ionisation), we only consider elastic collisions with the major constituent of the background gas, i.e. He atoms (which is justified by the low concentration of N$_2$ in the buffer gas): ${\rm He}^+ + {\rm He}  \rightarrow {\rm He}^+ + {\rm He}$ (which includes an isotopic channel and a backscattering channel), ${\rm N}_2^+ + {\rm He}  \rightarrow {\rm N}_2^+ + {\rm He}$, and ${\rm He}_2^+ + {\rm He}  \rightarrow {\rm He}_2^+ + {\rm He}$. For these processes we adopt the Langevin cross section:
\begin{equation}
\sigma_{\rm L} = \sqrt{\frac{\pi \alpha q^2}{\varepsilon_0 \mu}} \frac{1}{g},
\end{equation}
where $q$ is the elementary charge, $\alpha$ is the polarizibility of He atoms, $\mu$ is the reduced mass (of the projectile and target species), and $g$ is the relative velocity of the collision partners. The only additional heavy particle processes are the ${\rm He^*} + {\rm N}_2 \rightarrow {\rm He} + {\rm N}_2^+ + {\rm e}^-$ Penning ionisation and the ${\rm He}^+ + {\rm He} + {\rm He} \rightarrow {\rm He}_2^+ + {\rm He}$ ion conversion processes. The treatment of these two processes in the simulation proceeds as follows. The rates of these reactions (which are adopted from \cite{Brok,Sakiyama}) are used to assign a random lifetime -- according to the Monte Carlo approach -- to each of the metastable atoms and He$^+$ ions upon their "birth". These particles are then placed on a wait list and the given conversion reaction is executed at a later time according to the (random) lifetime of the given particle (for more details see \cite{Dns}). 

To account for the transport of the resonant radiation we keep track of the excited atoms in the He 2$^1$P state and trace the propagation of individual photons. The excited state is assigned to have a random lifetime $\tau_{\rm exc} = - \tau_0 \ln(1-R_{01})$ according to the natural lifetime of the 2$^1$P level, $\tau_0 = 0.56$ ns \cite{Zitnik}. Upon emission of the photon, natural broadening and pressure broadening, as well as Doppler broadening of the emission spectrum are considered and the actual wavelength of an individual photon is sampled randomly from this broadened spectrum (i.e. using a Monte Carlo approach \cite{Fierro}). The free path of the photon is defined by the (wavelength dependent) photo-absorption cross section of the 1$^1$S $\rightarrow$ 2$^1$P transition that is evaluated for the photon with a given wavelength and for potentially absorbing atoms having a random velocity sampled from the Maxwell-Boltzmann distribution at the given gas temperature, $T_{\rm g}$. At the high-pressure conditions considered here photons propagate via a sequence of free flight - absorption - emission events until they eventually escape the plasma, therefore it is important to consider the wavelength redistribution described above, for an accurate treatment of their propagation. For more details see \cite{Fierro,Dns}.

To optimise the runtime of the code, different time steps are used for the various species. Among the charged species, the most demanding constraint is posed on the time step of the electrons due to their high collisionality in the atmospheric-pressure background. Here, we use time step of $\Delta t_{\rm e} = 4.5 \times 10^{-14}$ s for the electrons. For the ions significantly longer time steps are allowed, we use the sub-cycling procedure for these species, with time steps of $\Delta t_{\rm He^+} = 10 \, \Delta t_{\rm e}$ and $\Delta t_{\rm He_2^+} = \Delta t_{\rm N_2^+} = 100 \,\Delta t_{\rm e}$. These time steps, as well as the spatial numerical grid satisfy the relevant stability criteria of the PIC scheme. For the resonance photons, on the other hand, the high photoabsorption cross section poses a significantly shorter time step, as compared to electrons. Thus, oppositely to ions: we use "super-cycling" for the photons, in each $\Delta t$-long tick of the simulation the photons are traced with $\Delta t_{\rm ph} = \Delta t_{\rm e} / 10000$, i.e. the photon time step is $\Delta t_{\rm ph} = 4.5 \times 10^{-18}$ s. While this time step seems prohibitively small for a simulation that needs to run to 10-100 ns, as the number of photons is significantly smaller than the number of charged particles, the simulation runs are feasible even at this  extremely small required time step.

Throughout this study, we adopt a gas temperature of $T_{\rm g}$ = 300 K and assume an electron reflection probability of 0.2 at the electrodes \cite{Kollath}. The coefficients of ion-induced emission of secondary electrons from the electrodes are set as $\gamma_{\rm He^+} =$ 0.15, $\gamma_{\rm He_2^+} =$ 0.1, $\gamma_{\rm N_2^+} =$ 0.05, while for the VUV photons we adopt the value $\gamma_{\rm ph} =$ 0.1 \cite{Dns}. Simulation runs start with seeding electrons, as well as N$_2^+$ and He$_2^+$ ions within the discharge gap with an equal number, with a cosine spatial distribution. These charged particles emulate the charges remaining from a previous pulse. We assume a repetitive operation of the system, which we are, however, not able to follow due to the very intensive computational needs that prohibit simulation of low duty cycle settings, with typically $\sim$ kHz repetition rates. 

Finally we comment on the possible roles of recombination processes and the photoionisation of N$_2$ molecules by the He resonant radiation in the discharges studied here. The recombination coefficients of He$_2^+$ and N$_2^+$ ions with electrons are, respectively, $k_1 = 8.9 \times 10^{-15} (T_{\rm e} / T_{\rm g})^{-3/2}$ m$^{3}$ s$^{-1}$ and $k_2 = 4.8 \times 10^{-13} (T_{\rm e} / T_{\rm g})^{-1/2}$ m$^{3}$ s$^{-1}$. These processes may become important only in the afterglow plasma. Our estimation, based on the fact that in the early afterglow $T_{\rm e} \gg T_{\rm g}$, indicates that these processes do not play a significant role for the conditions and time scales covered here. Photoionisation of N$_2$ molecules by the He resonant radiation is known to be an important process for some conditions, like for ambient air, i.e. at high N$_2$ concentration. According to \cite{PION}, the cross section of this reaction is 23.1 $\times$ 10$^{-18}$ cm$^{2}$, which results in a free path of photons that largely exceeds the electrode gap at the small N$_2$ concentration considered here. Thus this process can be safely neglected.

\subsection{Driving voltage pulse shapes}

As already mentioned in section 1, simulations will be conducted for different voltage pulse shapes, including single and double (unipolar and bipolar) voltage pulses with various durations and amplitudes. These voltage pulse shapes approximate those used in experiments on microsecond -- nanosecond discharges operated at sub-atmospheric -- atmospheric pressures. One has to note, however, that the pulse shapes in experiments are also a result of the interplay between the power supply, the electrical network and the plasma itself. Therefore, in experiments, usually more complicated pulse shapes are found, see e.g. \cite{Pai2009,Ibuka2007,Iza2009,Huang2015,Roettgen2016,Machala,Uwe,Song}.

\begin{figure}[ht!]
\begin{center}
\includegraphics[width=0.6\textwidth]{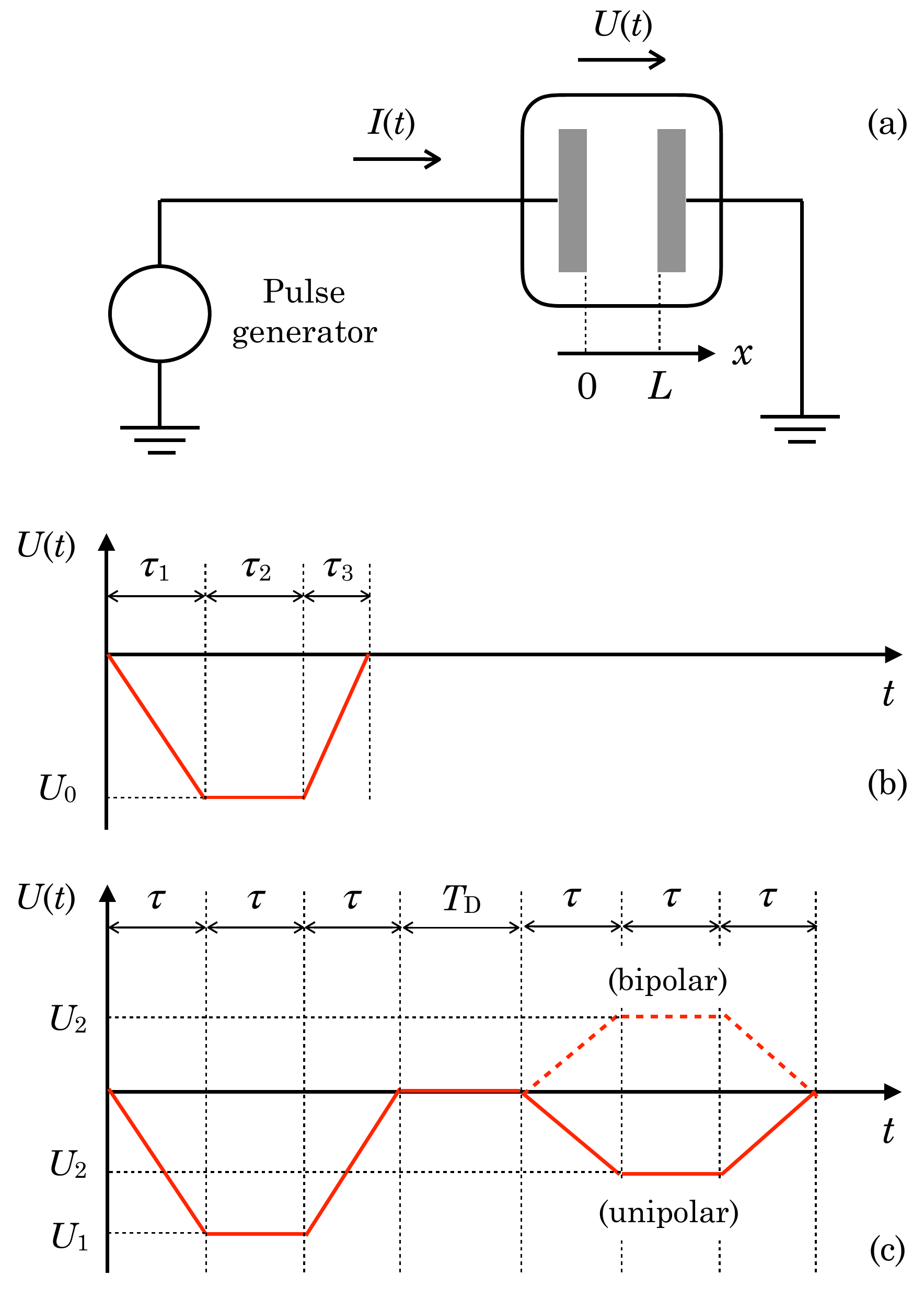}
\caption{Model electrical circuit (a) and excitation voltage pulse shapes: (b) Single-pulse waveform, with a  peak voltage $U_0$, rise, plateau duration, and fall times, $\tau_1$, $\tau_2$, and $\tau_3$, respectively. (Note that for single-pulse excitation $U_0$ is negative.) (c) Double-pulse waveforms: both pulses have the same rise, plateau duration, and fall times, $\tau$, and peak voltages, $U_1$ and $U_2$, respectively. $T_{\rm D} \geq 0$ is the delay time between the pulses. The continuous red line indicates a {\it unipolar} pulse ($U_1 < 0$, $U_2 < 0$), while the dashed line indicates the second pulse in the {\it bipolar} case ($U_1 < 0$, $U_2 > 0$).}
\label{fig:pulse-shapes}
\end{center}
\end{figure}

Figure~\ref{fig:pulse-shapes}(a) shows the model electrical circuit and the definition of the "measurements" of the discharge voltage, $U(t)$, and the discharge current, $I(t)$. The excitation voltage is applied at the electrode situated at $x$ = 0 mm, while the electrode at $x$ = 1 mm is grounded. Panels (b) and (c) of figure~\ref{fig:pulse-shapes} give a schematic representation of the pulse shapes, while table~\ref{table:cases} lists the cases presented in the forthcoming sections, with parameter values / ranges covered.  

% source: pulse-shapes.key

\begin{table}[h!]
\caption{\label{table:cases} Cases / effects studied in this work. All times ($\tau_1$, $\tau_2$, $\tau_3$, $\tau$, $T_{\rm D}$) in ns units, voltages ($U_0, U_1, U_2$) in Volts. The meaning of the parameters of the excitation voltage waveforms ($U(t)$) is defined in figure~\ref{fig:pulse-shapes}.}
\footnotesize
\begin{tabular}{@{}llllll}
\br
SINGLE-PULSE EXCITATION & $\tau_1$ & $\tau_2$ & $\tau_3$ & $U_0$ & figure(s)\\
\mr
general behavior & 5  & 5 & 5 & --1500 & \ref{fig:single-currents},~\ref{fig:s-meane},~\ref{fig:vuv-escape} \\
effect of pulse amplitude & 5  & 5 & 5 & --1000\dots--1800 & \ref{fig:single}(a) \\
effect of pulse width & 2\dots8  & 2\dots8 & 2\dots8 & --1500 & \ref{fig:single}(b) \\
effect of pulse ampl. \& width: "equivalent pulses"$^\ast$ & 5\dots12  & 5\dots12 & 5\dots12 & --1600 \dots --667 & \ref{fig:single}(c)\\
effect of pulse fall time & 5  & 5 & 1\dots 5 & --1500 &\ref{fig:pulse-fall-time} \\
post-pulse dynamics & 5  & 5 & 3 & --1500 &\ref{fig:post_maps},~\ref{fig:post_lines} \\
\br
DOUBLE-PULSE EXCITATION & $\tau$ & $T_{\rm D}$ & $U_1$  & $U_2$ & figure(s) \\
\mr
double unipolar vs. bipolar vs. single & 5 & 0 & --1300& $\mp$1300 , 0 & \ref{fig:comp3},~\ref{fig:comp3-ne}\\
double unipolar vs. "equivalent"$^\ast$ single & 5 & 0 & --1300& --1300 & \ref{fig:comp-shapes}\\
effect of delay time (unipolar case) & 3 & 0\dots50 & --1500 & --1500 & \ref{fig:delay}\\
effect of voltage ratio (bipolar case) & 5 & 0 & --1300 & 0\dots1300 & \ref{fig:U2}\\
effect of pulse amplitude (bipolar case) & 5 & 0 & --1000\dots--1700 & $-0.2\, U_1$ & \ref{fig:U1}\\
effect of pulse width (bipolar case) & 2\dots8 & 0 & --1500 & $-0.2\, U_1$ & \ref{fig:width}\\
\br
\end{tabular}\\
$^\ast$ "Equivalent" pulses are defined here as those having the same value of the $\int U(t) \,{\rm d}t$ integral.
\label{table:cases}
\end{table}

%\clearpage

\section{Results}

In the following we discuss the characteristics of the nanosecond discharges excited by the various waveforms introduced above. In section~\ref{sec:single}, we start with analysing the behaviour of the discharges driven by a single voltage pulse; here, following the illustration of the general characteristics of the discharges, we examine the effects of the voltage amplitude, the pulse length on the plasma characteristics, and address the effect of the falling slope of the voltage pulse on the post-pulse dynamics of the plasma. Subsequently, in section~\ref{sec:double}, we turn to the case of double-pulse excitation; we examine the cases of unipolar and bipolar driving pulses, address the questions how the voltage amplitude, the ratio of the amplitudes of the two pulses, the pulse duration, as well as the delay time between the two pulses influences the plasma characteristics. 

All the results presented here originate from simulations that include the VUV radiation transport and the important contribution of this radiation to the electron emission from the cathode. All results are presented for a gas mixture of He + 0.1\% N$_2$ at $p$ = 1 bar pressure,  $L$ = 1 mm electrode gap, $A$ = 1 cm$^2$ electrode area, a spatially averaged "seed" electron density $n_0 = 1.5 \times 10^{11}$ cm$^{-3}$, and a buffer gas temperature $T_{\rm g}$ = 300 K.

\subsection{Single-pulse excitation and post-pulse dynamics}

\label{sec:single}

\begin{figure}[h!]
\footnotesize{(a)}\includegraphics[width=0.43\textwidth]{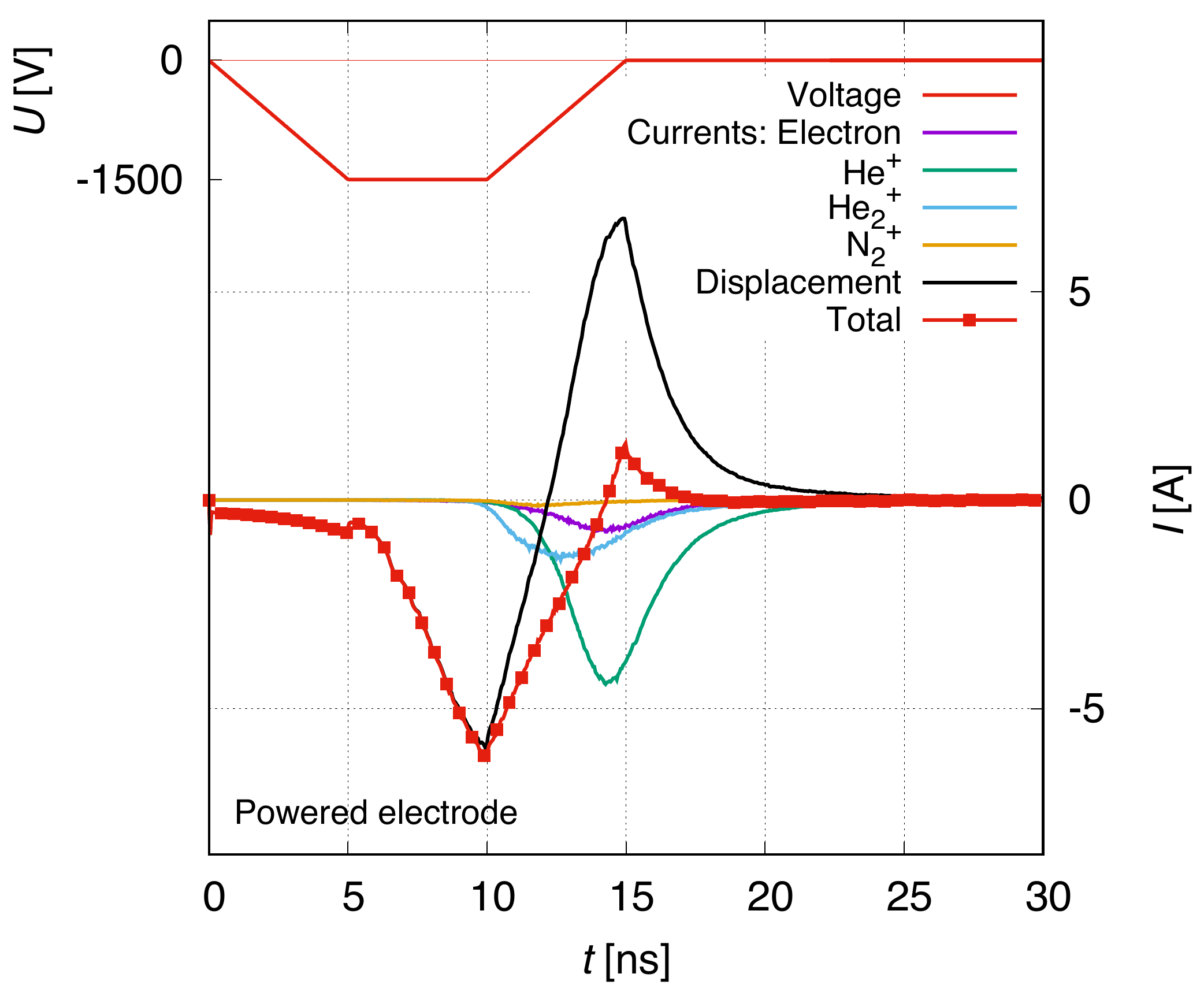}~~~~~~~
\footnotesize{(b)}\includegraphics[width=0.43\textwidth]{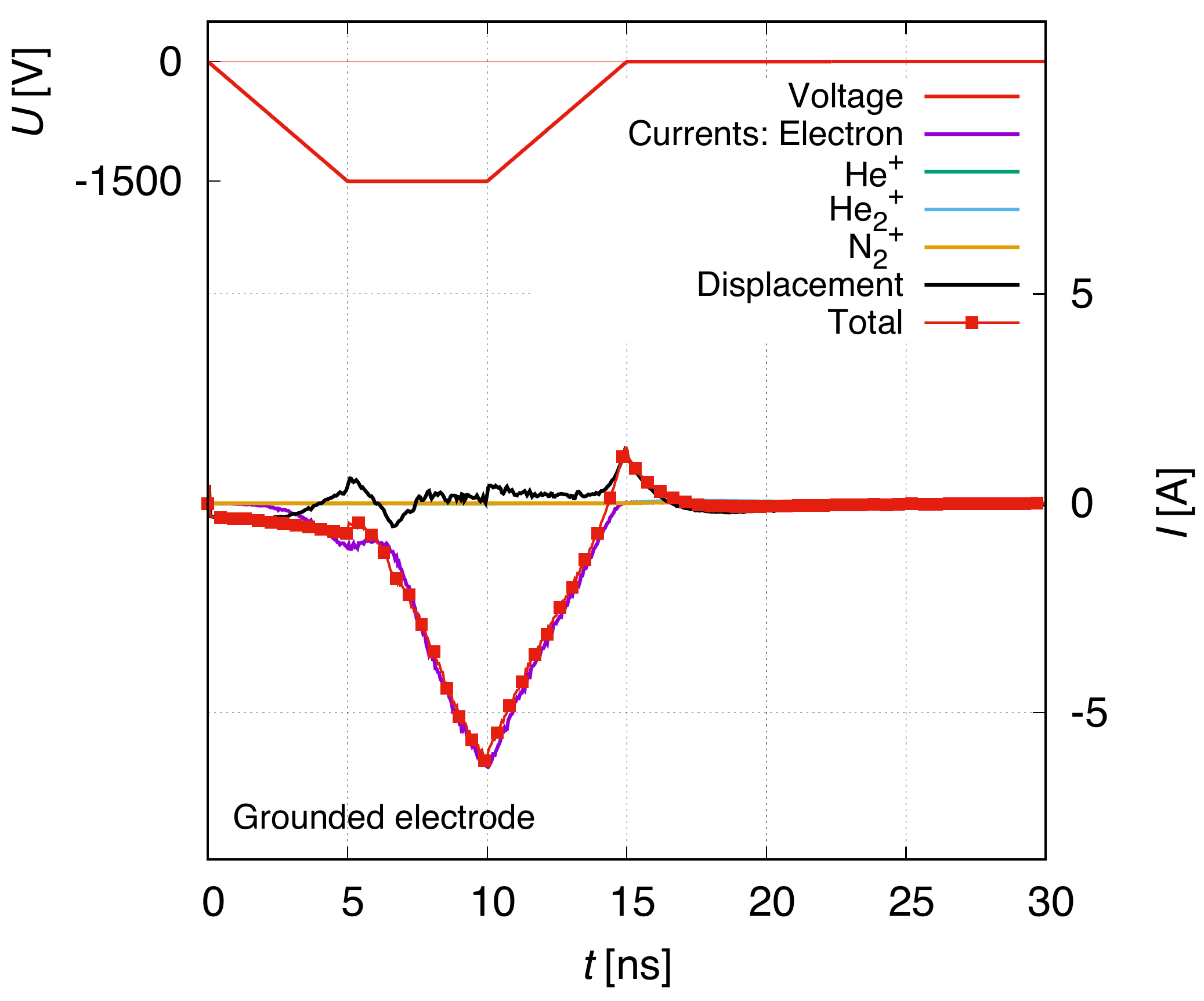}\\
\footnotesize{(c)}\includegraphics[width=0.48\textwidth]{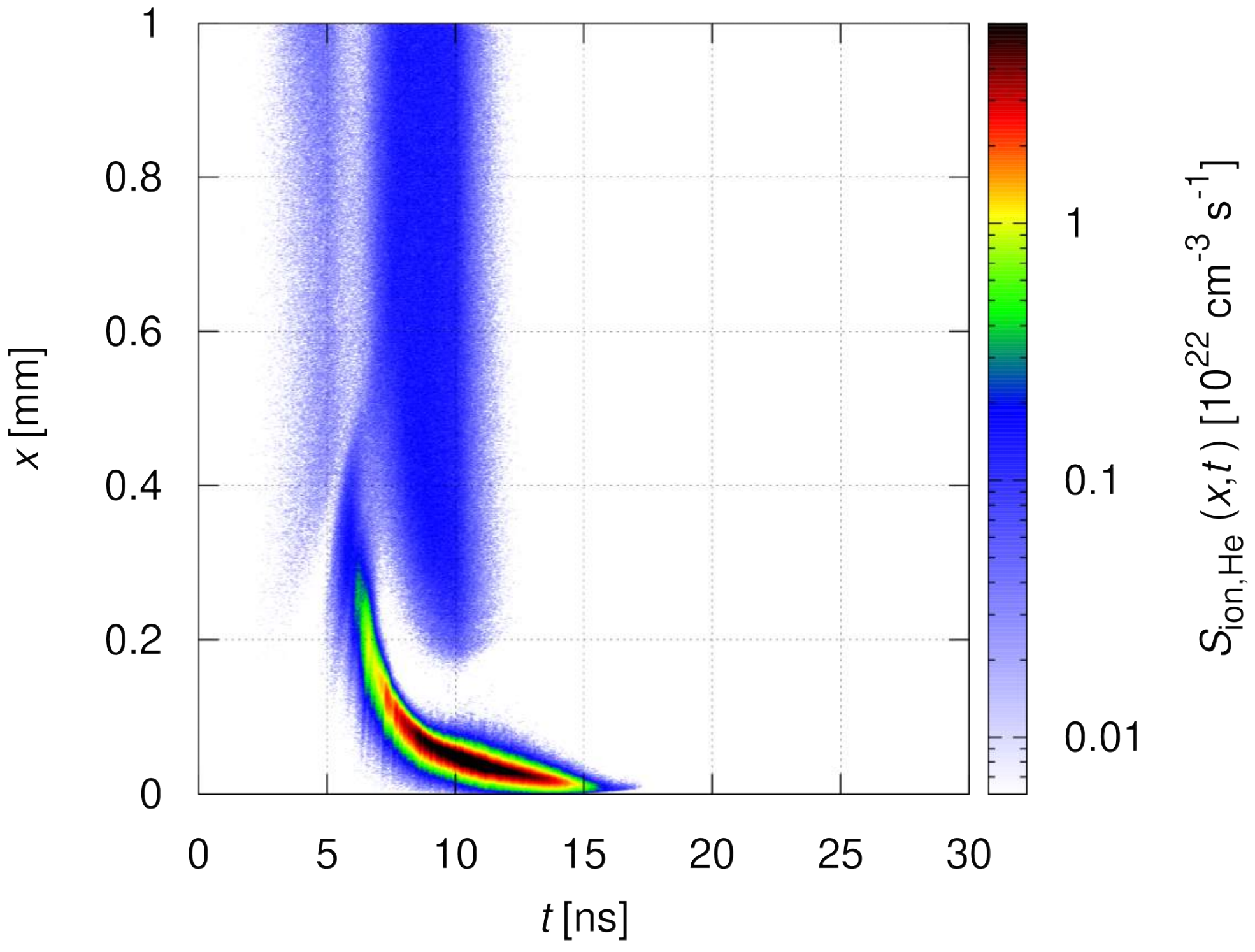}~
\footnotesize{(d)}\includegraphics[width=0.48\textwidth]{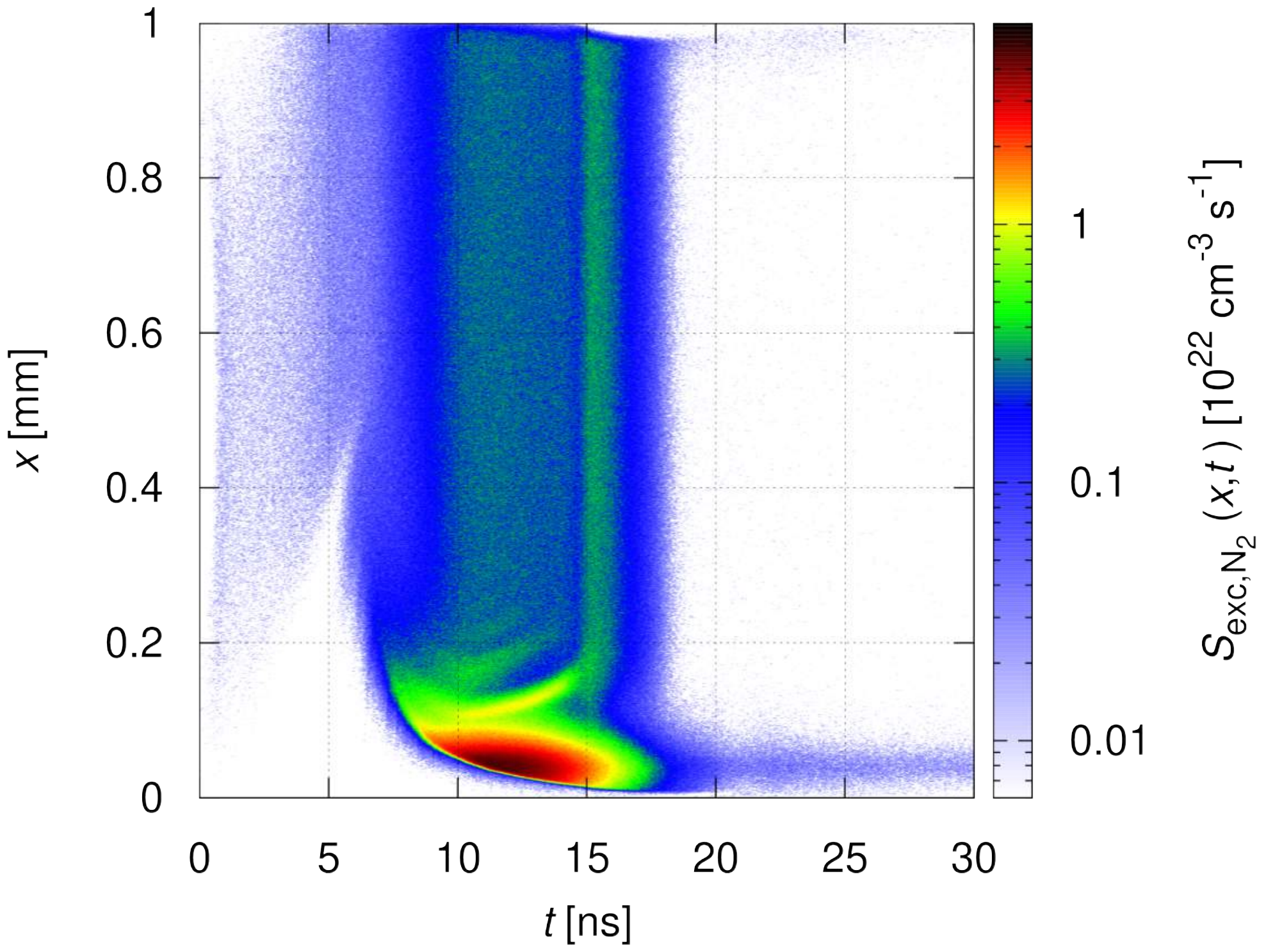}\\
\caption{Voltage pulse waveform (upper curves, left scales), as well as the total current and current components (lower sets of curves, right scales) at the powered electrode (a) and at the grounded electrode (b) of the discharge. Ionization rate of He (c) and total excitation rate of N$_2$ (d) as a function of position and time. Discharge conditions: $U_0 = -1500$ V and $\tau_1=\tau_2=\tau_3$ = 5 ns, He + 0.1\% N$_2$ at $p$ = 1 bar,  $L$ = 1 mm,  $n_0 = 1.5 \times 10^{11}$ cm$^{-3}$. The powered electrode is at $x$ = 0 mm, while the grounded electrode is at $x$ = 1 mm.}
\label{fig:single-currents}
\end{figure}
%source: single/single-currents-v2.plot
%source: single/single-maps.plot

Figure~\ref{fig:single-currents} illustrates the "general" behaviour of the discharge. Panels (a) and (b) display the discharge current and its components at the powered electrode (cathode, $x= 0$ mm) and at the grounded electrode (anode, $x=$ 1 mm), respectively. In panels (c) and (d) we show the spatio-temporal distributions of the ionisation rate of He atoms and the total excitation rate of N$_2$ molecules. The data were obtained for a voltage pulse amplitude of $U_0 = -1500$ V, and $\tau_1=\tau_2=\tau_3$ = 5 ns. The current components include the conduction currents of the different charged species, as well as the displacement current induced by the temporal variation of the electric field. 

At the powered electrode of the discharge, as seen in figure~\ref{fig:single-currents}(a), only displacement current ($I_{\rm d}$) flows at times $t \lesssim$ 10 ns, as the number of ions that reach the electrode within this time period is very small. During the rising slope of the pulse the magnitude of $I_{\rm d}$ increases slowly, because the electric field is nearly homogeneous at these times. Subsequently, when space charges start to modify the electric field distribution remarkably, the magnitude of the electric field in the vicinity of this electrode increases rapidly, giving rise to a rapidly rising $|I_{\rm d}|$. At $t >$ 10 ns the magnitudes of the voltage and the electric field start to decrease, and $I_{\rm d}$ changes sign. Significant ion currents to this electrode flow only during the falling slope of the voltage pulse, at the given conditions. Among the ion currents, that of the He$^+$ ions is dominant. The ion currents induce as well an electron current from the negatively biased powered electrode, with an order of $-1$ A peak value. The currents at the grounded (anode) side of the discharge (see figure~\ref{fig:single-currents}(b)) are completely different. There, the electron current dominates at all times, the displacement current gives a small contribution, while positive ion currents are negligible (although non-zero after the termination of the voltage pulse, during the decay of the plasma (not resolved in the figure)). The characteristic positive ("reverse") peak of the current upon the termination of the voltage pulse will be discussed later in this section.

The evolution of the discharge, in space and time, is well reflected by the He ionisation rate plotted in figure~\ref{fig:single-currents}(c). During the rise time of the voltage pulse (first 5 nanoseconds) the ionisation rate grows slowly, He$^+$ ions during this time are created by the "seed" electrons that are accelerated towards the anode. Around $t$ = 6 ns an ionisation peak forms near the centre of the electrode gap, which moves to the vicinity of the cathode during the next few nanoseconds. The ionisation rate is maximum in this domain, at a value of $\sim 6 \times 10^{22}$ cm$^{-3}$ s$^{-1}$. Meanwhile a bulk plasma forms and gradually fills most of the gap. The total excitation rate of N$_2$ molecules (figure~\ref{fig:single-currents}(d)) is more spread as compared to the ionisation rate of He, due to the fact that processes with significantly lower excitation thresholds (than that of He$^+$ formation) are included in this distribution. Excitation of N$_2$ molecules is seen even after the termination of the voltage pulse, which is related to the post-pulse plasma dynamics, where lower threshold (rotational and vibrational) levels can still be excited thanks to the electric field created within the plasma. 

The striated structure that appears at the cathode side of the column region in figure~\ref{fig:single-currents}(d) is analogous to that usually observed at low pressures (striated positive column). Such structures are typical for conditions where the electron energy relaxation proceeds in a periodic manner (via repeating energy gain - collisional energy loss cycles), at intermediate values of the reduced electric field, $E/N$ \cite{periodic1,periodic2,periodic3,periodic4,periodic5}. The reduced electric field for the case studied in figure~\ref{fig:single-currents} is a few tens of Townsends (1 Td = 10$^{-21}$ V\,m$^2$) in the column region, which belongs to the favourable range of $E/N$ for striations. We note that periodic structures were also found in simulations of plasma jets operated with radio-frequency excitation at similar values of the gap length and gas pressure. There, however, these structures formed in the vicinity of the electrodes, in regions with "proper" $E/N$ values  \cite{Bochum_paper}.    

\begin{figure}[ht!]
\begin{center}
\includegraphics[width=0.5\textwidth]{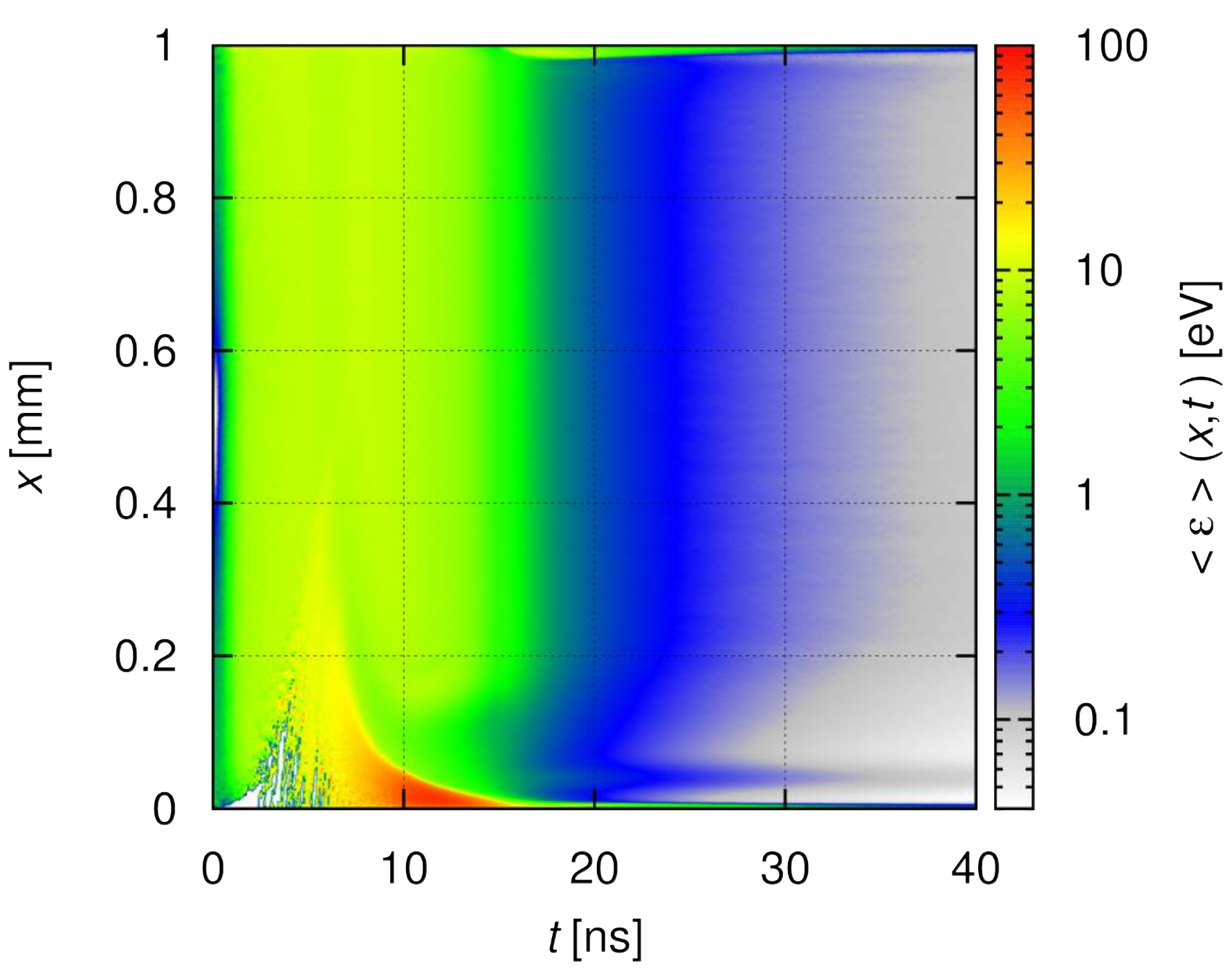}
\caption{Time and space dependence of the mean energy of electrons. ($U_0 = -1500$ V, $\tau_1=\tau_2=\tau_3$ = 5 ns, He + 0.1\% N$_2$ at $p$ = 1 bar,  $L$ = 1 mm,  $n_0 = 1.5 \times 10^{11}$ cm$^{-3}$. The powered electrode is at $x$ = 0 mm, while the grounded electrode is at $x$ = 1 mm.)}
\label{fig:s-meane}
\end{center}
\end{figure}
%source: single/single-meane.plot

\begin{figure}[ht!]
\begin{center}
\includegraphics[width=0.5\textwidth]{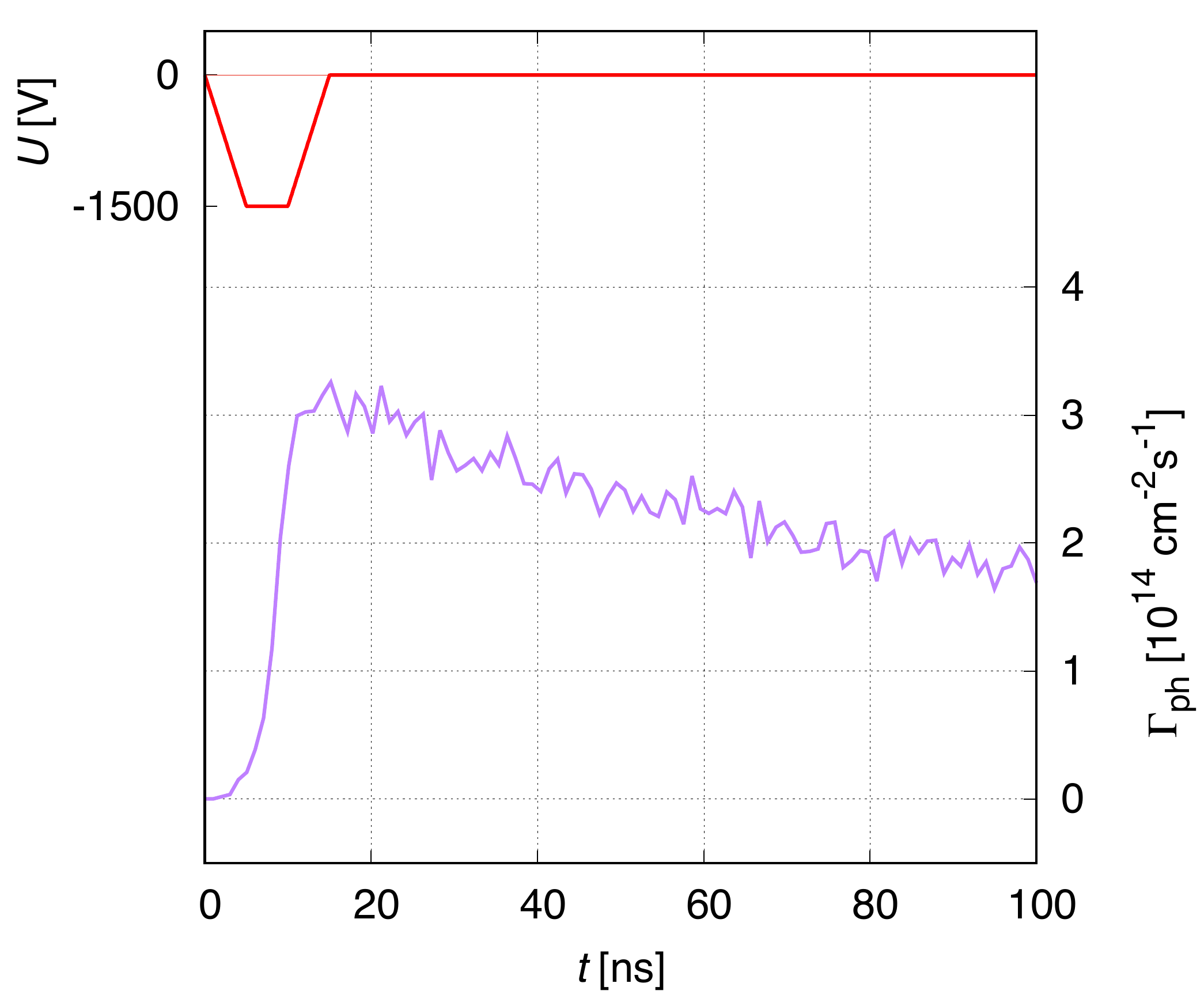}
\caption{Flux of the VUV helium resonance radiation at the cathode (lower curve, right scale) obtained in the case of an excitation waveform $U_0 = -1500$ V and $\tau_1=\tau_2=\tau_3$ = 5 ns (upper curve, left scale). Note that the radiation persists at times well after the excitation pulse. (Other conditions: He + 0.1\% N$_2$ at $p$ = 1 bar,  $L$ = 1 mm,  $n_0 = 1.5 \times 10^{11}$ cm$^{-3}$.)}
\label{fig:vuv-escape}
\end{center}
\end{figure}
%source: single/photons-v2.plot

The mean electron energy, $\langle \varepsilon \rangle$, is plotted in figure~\ref{fig:s-meane}. At early times, during the breakdown phase the mean energy acquires a value in the order of 10 eV in the whole discharge gap. When the sheath forms $\langle \varepsilon \rangle$ grows to several times this value, within a narrow region close to the powered electrode (cathode), as the consequence of the very high electric field in that region that accelerates electrons emitted from the cathode to very high energies. The mean energy in the column region (during the active phase of the discharge) remains $\sim$ 10 eV. Following the termination (at $t=$ 15 ns) of the excitation, $\langle \varepsilon \rangle$ starts to decay, however, the values between $t=$ 20 ns and 30 ns indicate the presence of a significant electric field. At later times the electrons thermalise with the background gas, however, this process is not described correctly by our simulations because we adopted the cold-gas approximation for the electrons, the consequence of which is $\langle \varepsilon \rangle \rightarrow 0$, instead of the physically correct $\langle \varepsilon \rangle \rightarrow \frac{3}{2} k_{\rm B} T_{\rm g}$. Within the narrow sheath regions adjacent to both electrodes at $t >$ 15 ns, $\langle \varepsilon \rangle$ amounts several eV, which is due to the acceleration of electrons emitted from both electrodes due to ion and photon impact, in the ambipolar fields in these regions (see later).

We note that there is a significant energy storage in the plasma even after the termination of the voltage pulse - a high number of metastable atoms and excited atoms in the 2$^1$P resonant state are present. An evidence of the latter is the slowly decaying intensity of the resonant VUV photon flux at the electrodes, shown in figure~\ref{fig:vuv-escape} (for $U_0 = -1500$ V and $\tau_1=\tau_2=\tau_3$ = 5 ns). The slow decay of this radiation is explained by the fact that under the high-pressure conditions the plasma is optically thick for the VUV resonance radiation, therefore photons escape from the plasma on time scales that can be substantially longer ($\sim \mu$s) than the duration of the excitation pulses. 

% end of presentation of general characteristics

\begin{figure}[ht!]
\begin{center}
\footnotesize{(a)}\includegraphics[width=0.45\textwidth]{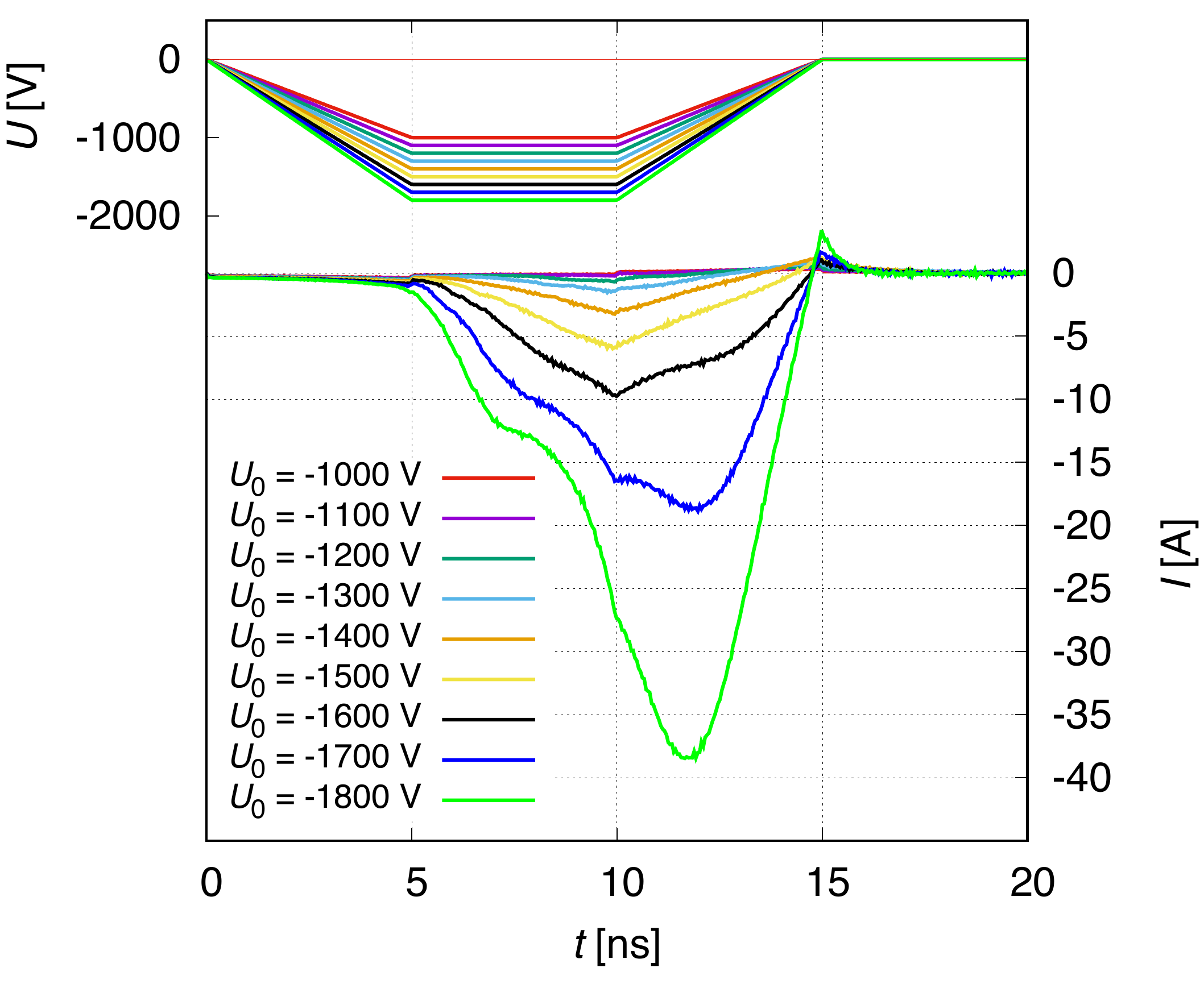}\\
\footnotesize{(b)}\includegraphics[width=0.45\textwidth]{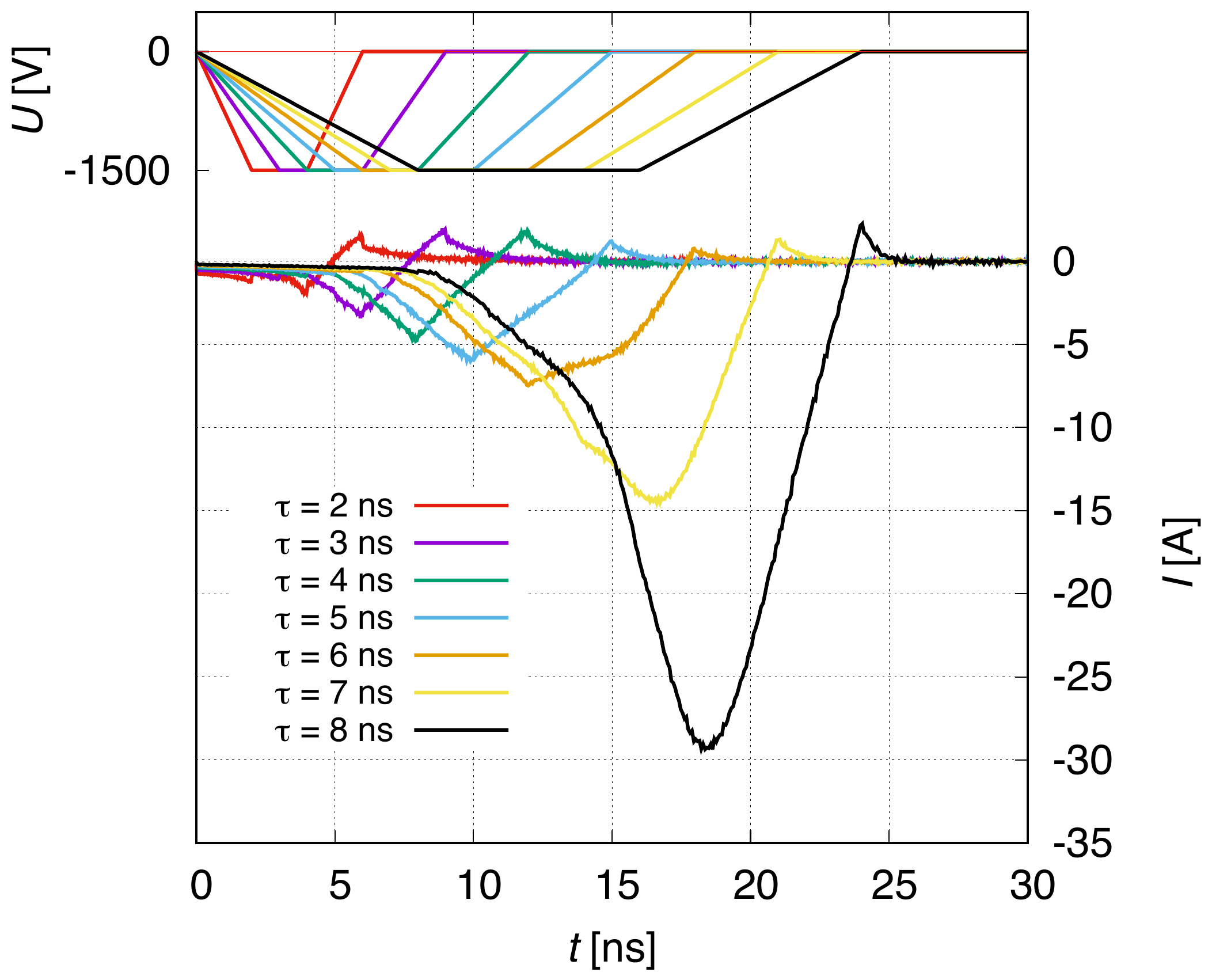}\\
\footnotesize{(c)}\includegraphics[width=0.45\textwidth]{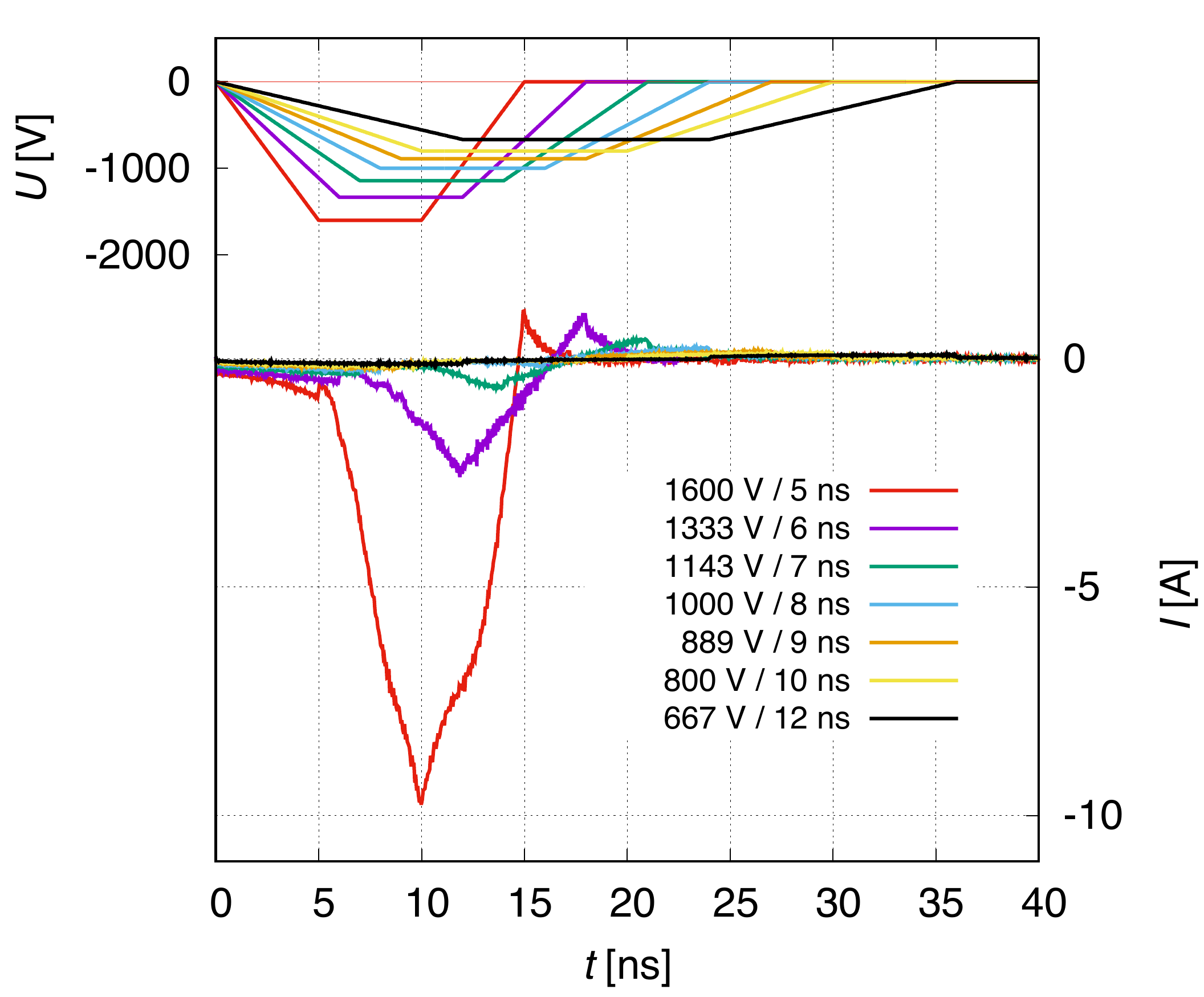}
\caption{Effects of changing the voltage pulse amplitude $U_0$ at fixed pulse duration $\tau$ = 5 ns (a), the pulse duration $\tau$ at fixed $U_0$ = $-1500$ V(b), and both of these parameters but keeping their product constant at $U_0 \, \tau$ = -8000 V ns (c), on the discharge current pulse. The upper and lower sets of curves (in each panel) correspond, respectively, to the driving voltage (left scale) and the computed current (right scale). (Other conditions: He + 0.1\% N$_2$ at $p$ = 1 bar,  $L$ = 1 mm,  $n_0 = 1.5 \times 10^{11}$ cm$^{-3}$.)}
\label{fig:single}
\end{center}
\end{figure}
%source: single/single-effects-v2.plot
%source: single/single-strech.plot

For the case of single-pulse excitation, the effect of the voltage amplitude on the computed discharge current is illustrated in figure~\ref{fig:single}(a). The excitation voltage pulse has the same rise, plateau duration and fall times, $\tau_1=\tau_2=\tau_3 = \tau$ = 5 ns, its amplitude is varied between $U_0 = -1000$ V and $-1800$ V. Figure~\ref{fig:single}(b) shows the effect of a varying pulse duration, at a fixed amplitude $U_0 = -1500$ V (for $\tau_1=\tau_2=\tau_3$). Both parameters have drastic effects on the shape and peak value of the current pulses. A higher voltage increases electron multiplication significantly during the breakdown phase, creating a higher density plasma, with higher conductivity. Consequently, the peak current of the pulse scales strongly non-linearly with the voltage amplitude (see figure~\ref{fig:single}(a)). The scan of the pulse duration at fixed peak voltage, shown in figure~\ref{fig:single}(b) indicates that at the peak voltage, $U_0 = -1500$ V chosen, the plasma development is far from being complete when an excitation pulse with $\tau$ = 8 ns is applied, and that the current would grow to exceedingly high values if longer excitation pulses are applied. In other words, this study confirms that the plasma characteristics, e.g., charged particle densities, in the parameter domain studied here are limited to a large extent by the length of the voltage pulse. An additional scan over the parameters is presented in figure~\ref{fig:single}(c), where both $U_0$ and $\tau$ are changed, keeping, however, their product, i.e. the integral $\int U(t) {\rm d}t$ fixed, at $U_0 \, \tau$ = -8000 V\,ns. A significant decay of the peak current is observed as a result of the "stretching" of the excitation pulse.

\begin{figure}[ht!]
\begin{center}
\includegraphics[width=0.48\textwidth]{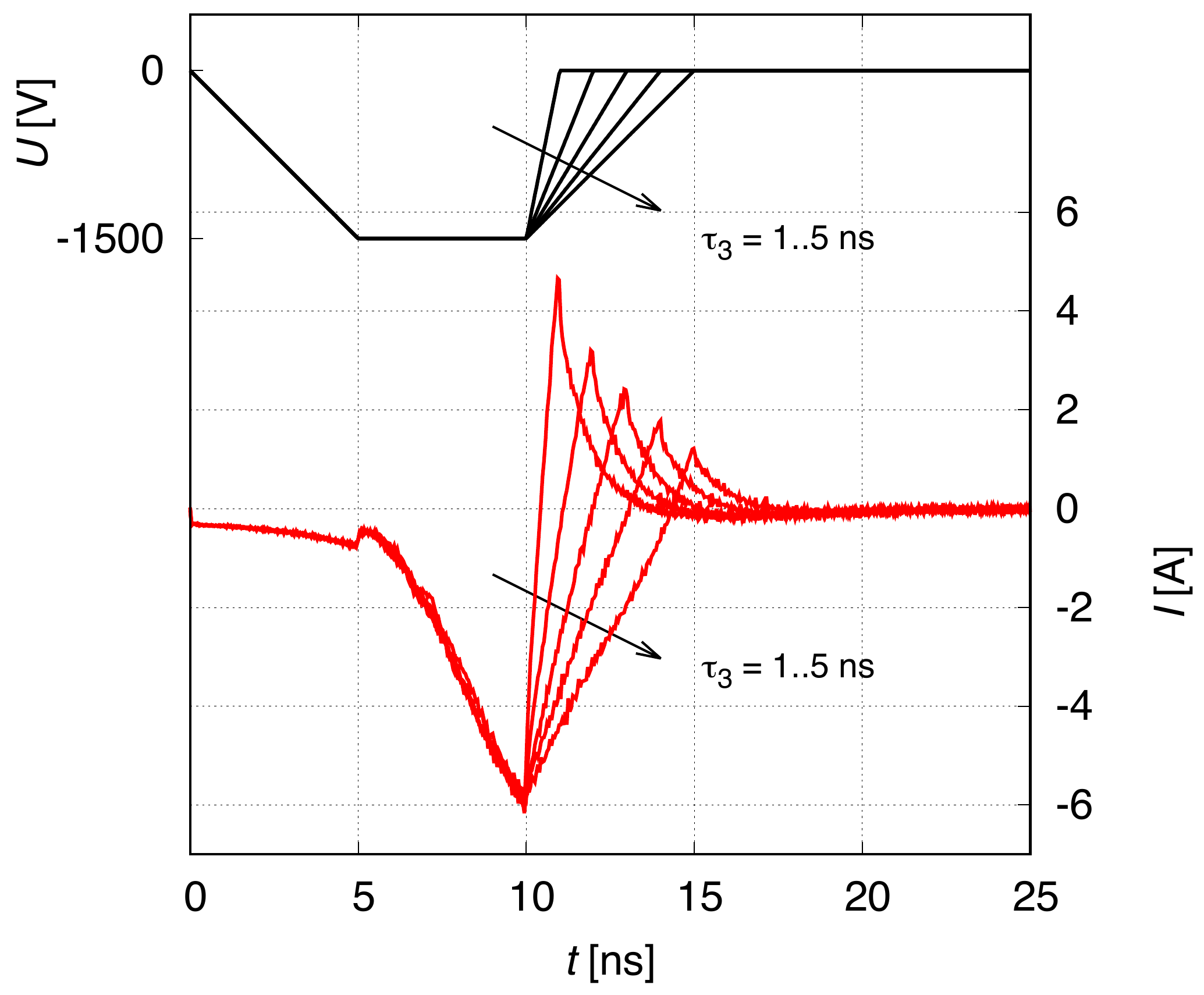}
\caption{The effect of the fall-time of the excitation voltage pulse (upper curves, left scale), $\tau_3$, on the discharge current (lower curves, right scale) for $U_0 = -1500$ V and $\tau_1$ = $\tau_2$ = 5 ns. (Other conditions: He + 0.1\% N$_2$ at $p$ = 1 bar,  $L$ = 1 mm,  $n_0 = 1.5 \times 10^{11}$ cm$^{-3}$.)}
\label{fig:pulse-fall-time}
\end{center}
\end{figure}
% source: post_pulse/postpulse-currents-corrected-v2.plot

\begin{figure}[ht!]
\begin{center}
\footnotesize{(a)}\includegraphics[width=0.45\textwidth]{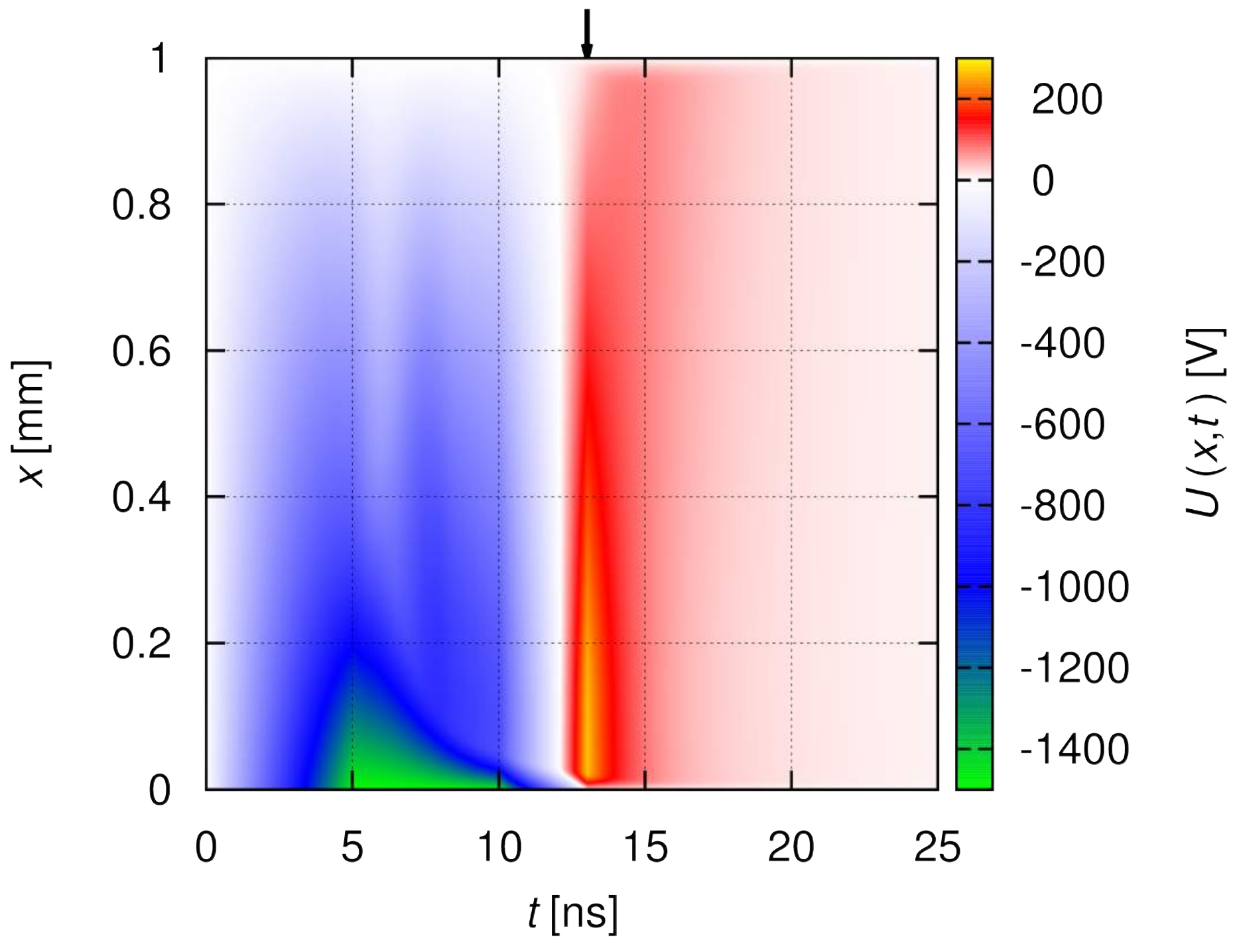}
\footnotesize{(b)}~~\includegraphics[width=0.45\textwidth]{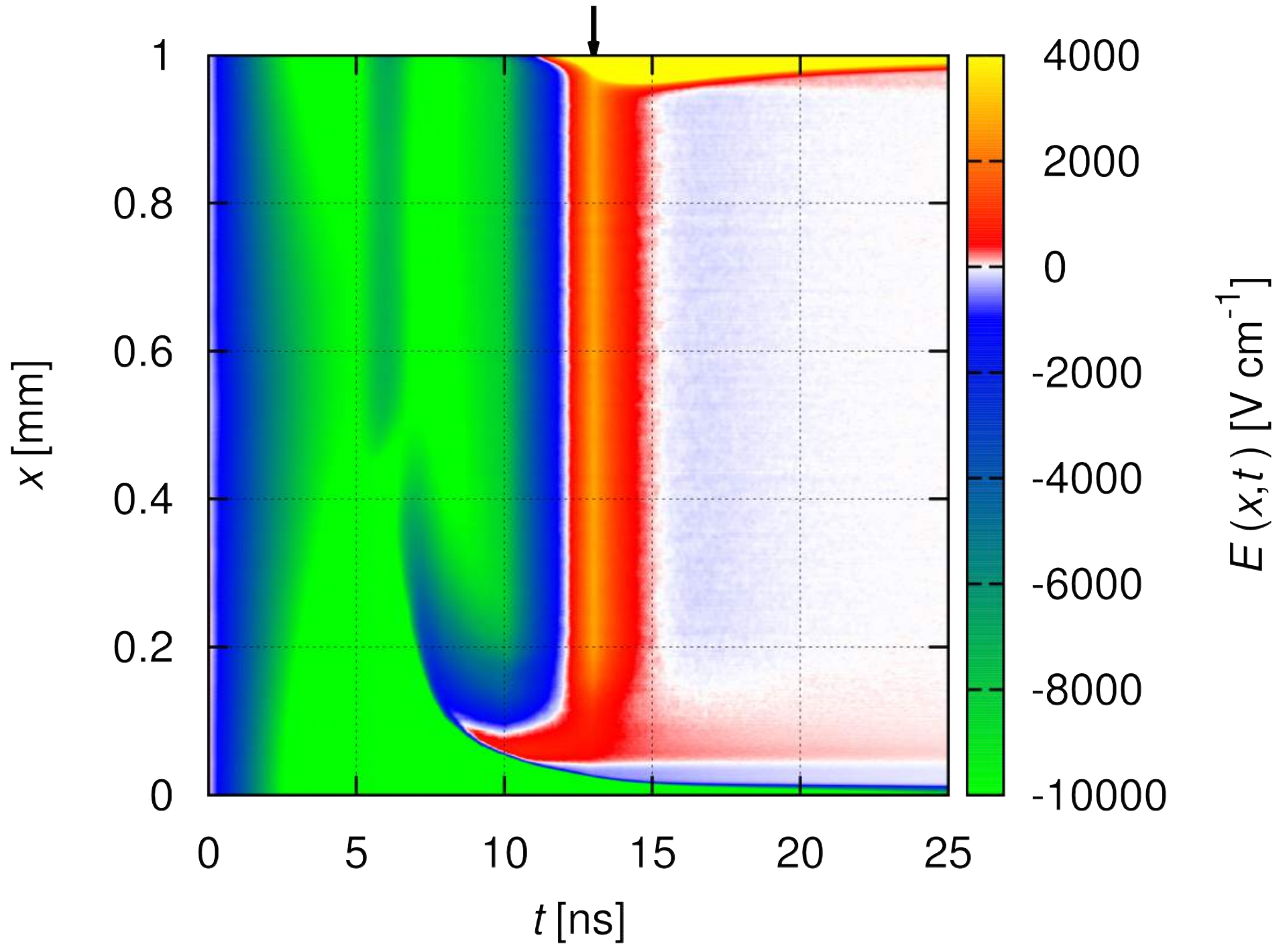}\\
\footnotesize{(c)}\includegraphics[width=0.45\textwidth]{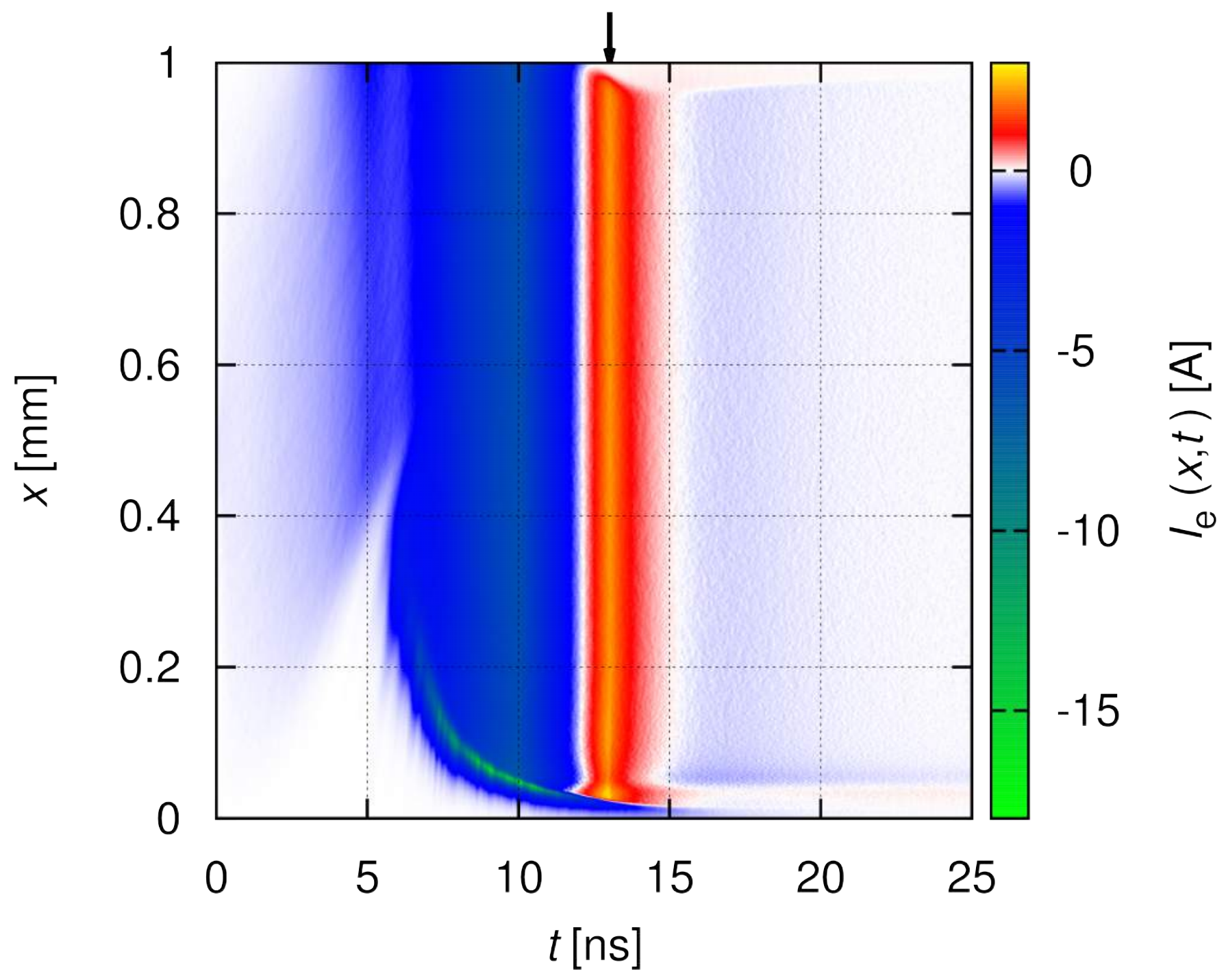}
\footnotesize{(d)}~~\includegraphics[width=0.45\textwidth]{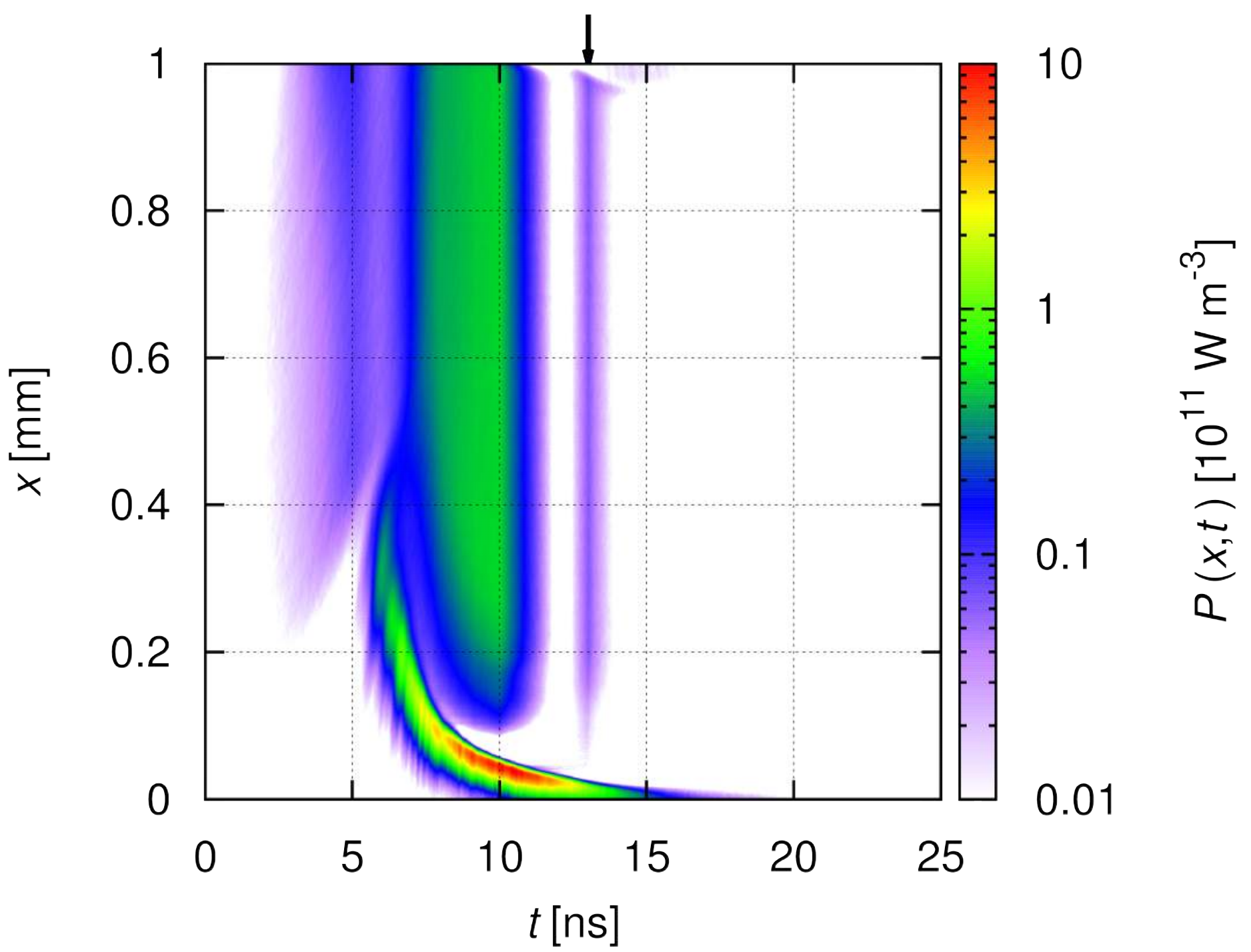}
\caption{Spatio-temporal distributions of the (a) potential, (b) electric field, (c) conduction current, and (d) power density absorbed by the electrons. Discharge conditions: $U_0 = -1500$ V,  $\tau_1 = \tau_2$ = 5 ns, $\tau_3$ = 3 ns. The colour scale in (b) is saturated in the high-field regions of the discharge. The black arrows at the top of figures mark the termination of the excitation voltage pulse ($t$ = 13 ns). (Other conditions: He + 0.1\% N$_2$ at $p$ = 1 bar,  $L$ = 1 mm,  $n_0 = 1.5 \times 10^{11}$ cm$^{-3}$. The powered electrode is at $x$ = 0 mm, while the grounded electrode is at $x$ = 1 mm.)}
\label{fig:post_maps}
\end{center}
\end{figure}

\begin{figure}[ht!]
\begin{center}
\footnotesize{(a)}\includegraphics[width=0.44\textwidth]{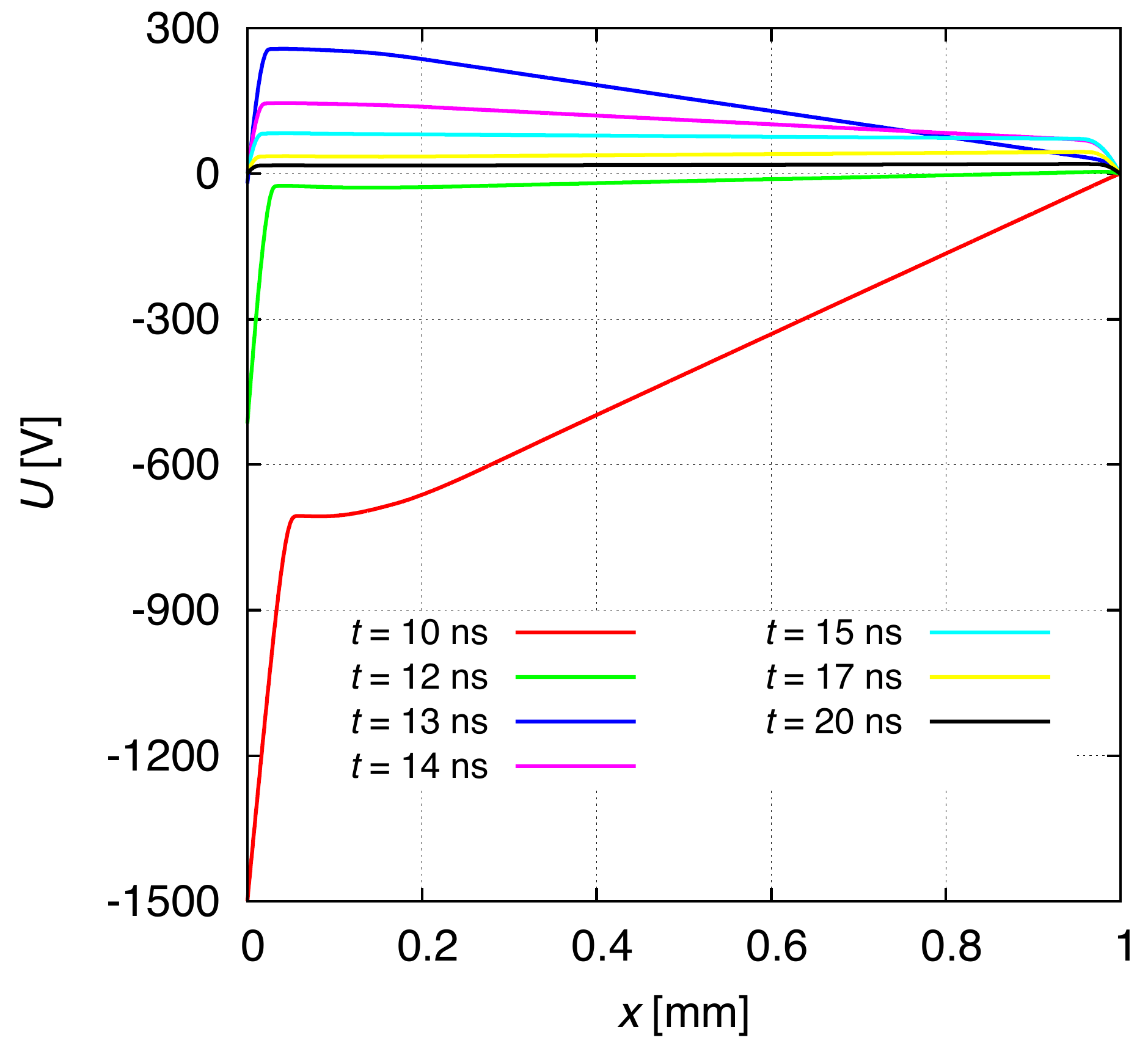}
\footnotesize{(b)}\includegraphics[width=0.44\textwidth]{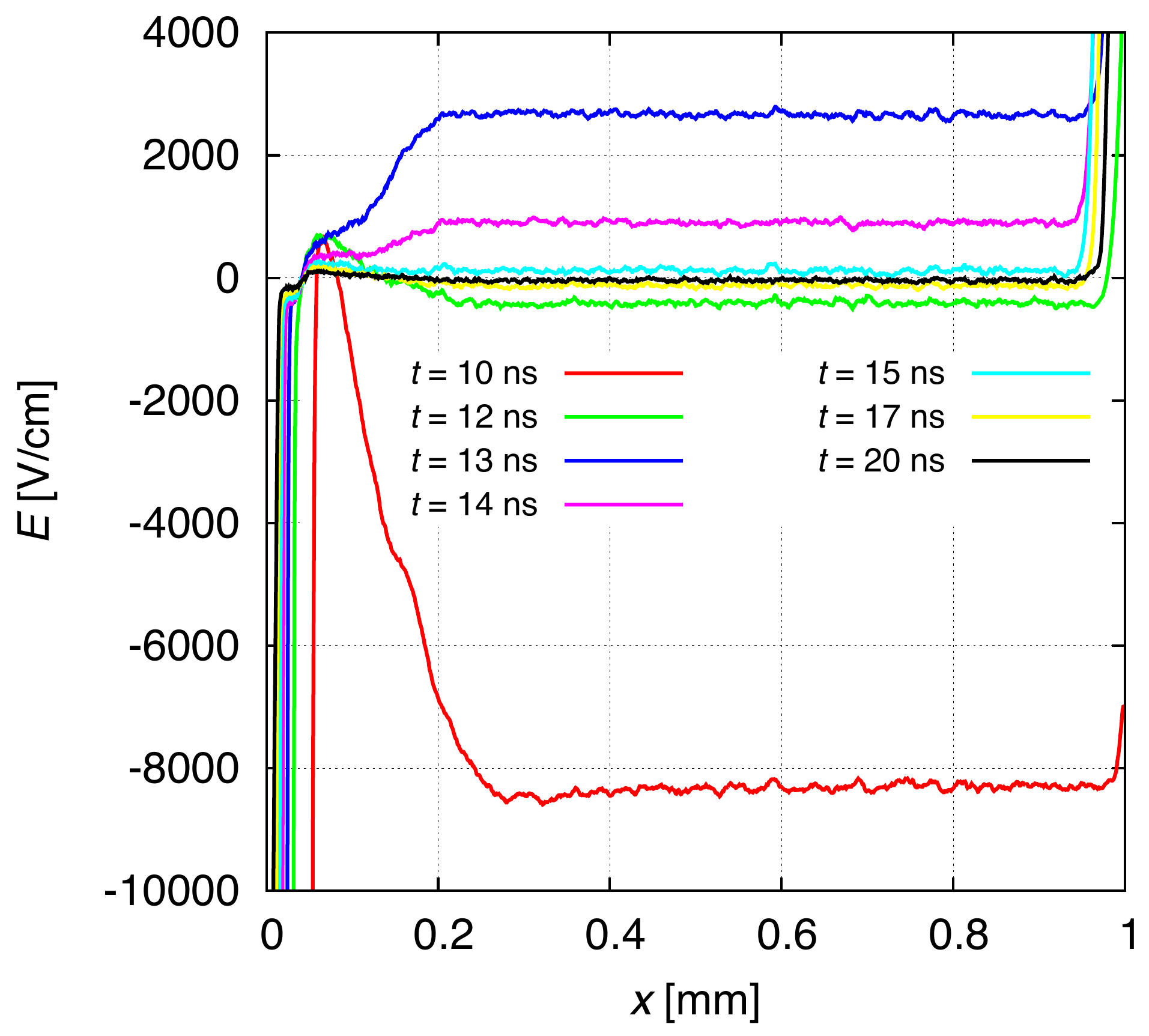}\\
\footnotesize{(c)}~~~~~\includegraphics[width=0.44\textwidth]{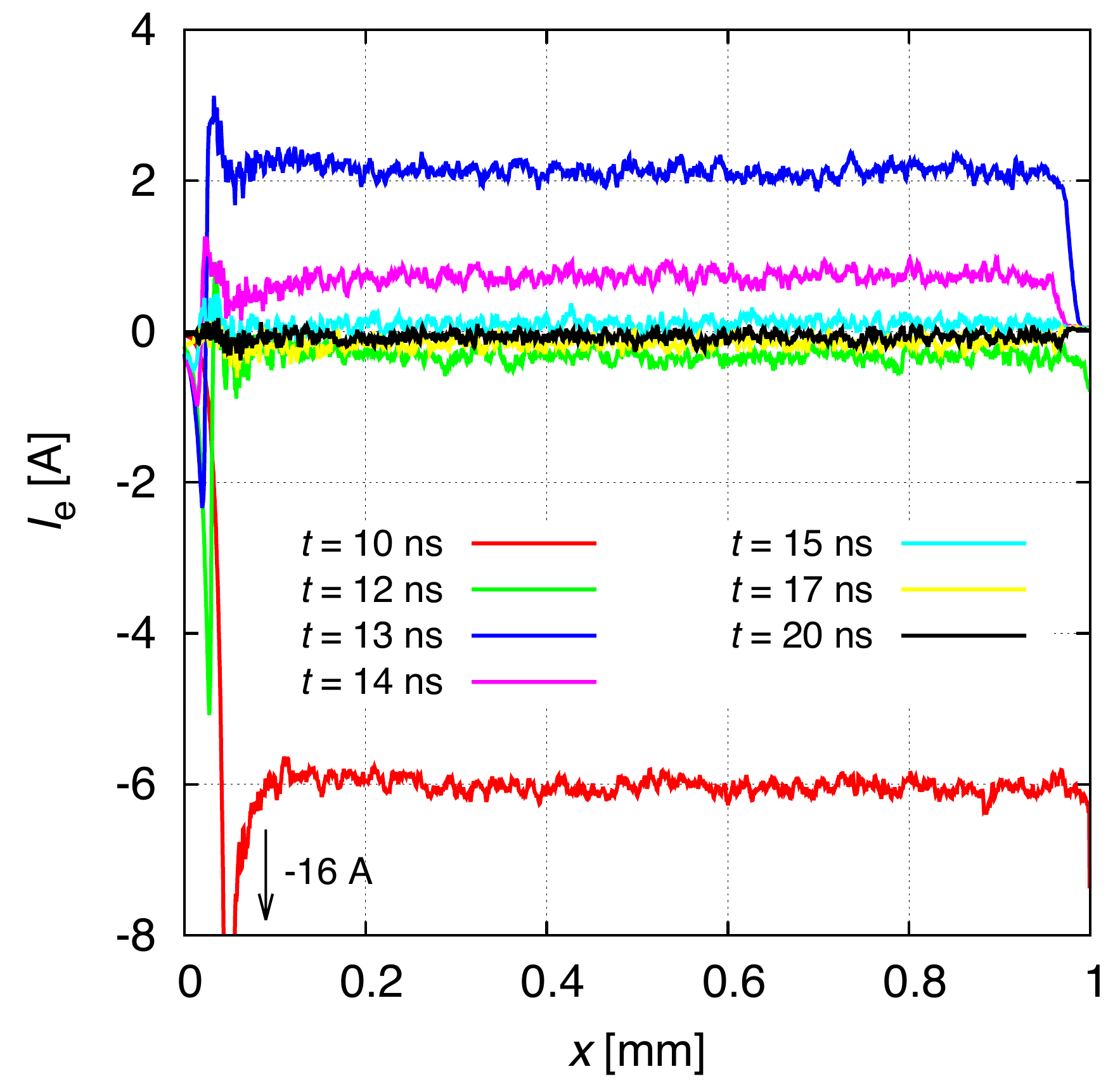}~~~~~~~~
\footnotesize{(d)}\includegraphics[width=0.42\textwidth]{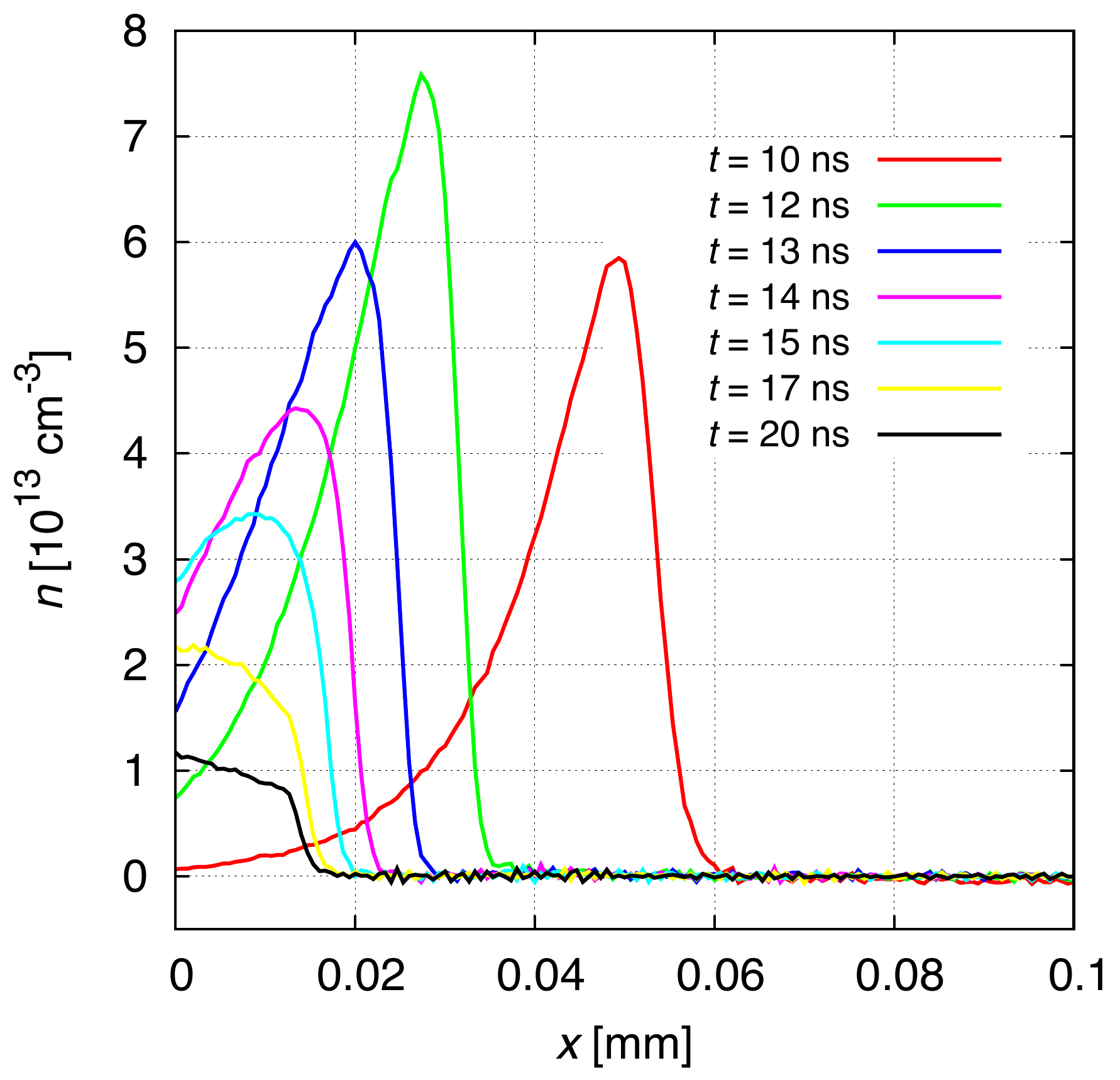}
\caption{(a) Potential, (b) electric field, (c) electron current, and (d) net charged particle density (ion density minus electron density) in the plasma at selected times. Discharge conditions: $U_0 = -$ 1500 V,  $\tau_1 = \tau_2$ = 5 ns, $\tau_3$ = 3 ns. The large values of the electric field near the electrodes are out of scale of panel (b). The net density in (d) is shown only for the vicinity of the cathode. (Other conditions: He + 0.1\% N$_2$ at $p$ = 1 bar,  $L$ = 1 mm,  $n_0 = 1.5 \times 10^{11}$ cm$^{-3}$. The powered electrode is at $x$ = 0 mm, while the grounded electrode is at $x$ = 1 mm.)}
\label{fig:post_lines}
\end{center}
\end{figure}

The formation of a "reverse" (positive) current peak upon the termination of the voltage pulse is observed for all conditions. This feature is, in fact, the primary macroscopic manifestation of the post-pulse dynamics. The appearance of this peak was already noted in, e.g., \cite{Dns} and in figure~\ref{fig:single-currents} above. Figure~\ref{fig:pulse-fall-time} reveals that the magnitude of this current peak exhibits a strong dependence on the fall time ($\tau_3$) of the excitation voltage pulse. (The cases shown correspond to $U_0 = -1500$ V and $\tau_1=\tau_2$ = 5 ns.) As $\tau_3$ decreases from 5 ns to 1 ns, we observe an almost five times increase of the magnitude of the reverse  current peak. The effects generating this current pulse, and the post-pulse dynamics of the plasma are analysed with the aid of the spatio-temporal distributions of selected discharge characteristics. The potential, the electric field, the electron conduction current, and the power density absorbed by the electrons are depicted in figure~\ref{fig:post_maps}. Additional plots, showing the spatial distributions of some of these characteristics, as well as that of the net (i.e. the positive $-$ negative) particle density are shown in figure \ref{fig:post_lines}, at selected values of time. For this analysis we choose the case with $U_0 = -1500$ V and $\tau_1=\tau_2$ = 5 ns, $\tau_3$ = 3 ns. 

% source: post_pulse/post-pulse-maps.plot
% source: post_pulse/post-pulse-eheat.plot

% source: post_pulse/post-pulse-lines.plot

The spatio-temporal potential distribution, $U(x,t)$, (figure~\ref{fig:post_maps}(a)) during the duration of the pulse shows a large voltage drop near the cathode (i.e. the cathode fall voltage) while the remaining voltage drops over the column region. This distribution is also well seen in the "$t$ = 10 ns" graph of figure~\ref{fig:post_lines}(a): about 800 V drops near the cathode (over $\approx$ 0.05 mm), then a field-free region follows, and the rest of the voltage, 700 V at this time, drops over the column region spanning from $x \approx$ 0.2 mm to $x$ = 1 mm. During the falling edge of the excitation voltage pulse (10 ns $\leq t \leq$ 13 ns) the potential over the whole gap decreases, however, right at the termination of the pulse (at $t$ = 13 ns) a positive plasma potential is generated that peaks at $\approx$ 250 V. This potential is generated by the net positive space charge, see  figure~\ref{fig:post_lines}(d). The electric field also changes sign upon the termination of the excitation (figures~\ref{fig:post_maps}(b)) and \ref{fig:post_lines}(b)). At the anode side ($x$ = 1 mm), similarly to the cathode side ($x$ = 0 mm), a space charge sheath also develops after the voltage pulse - these two sheaths generate ambipolar fields to confine electrons within the afterglow plasma. The corresponding potential distribution, e.g. at $t$ = 15 ns, is specific for such case (see figure~\ref{fig:post_lines}(a)), the strong ambipolar fields are also well seen in figure~\ref{fig:post_maps}(b) near the electrodes, at $t >$ 13 ns.

The electric field changes sign during the falling edge of the voltage pulse and reaches its highest positive value within the bulk upon the termination of the voltage pulse. Following the change of direction of $E$, the electron current changes sign instantaneously, as it can be seen in figures~\ref{fig:post_maps}(c) and \ref{fig:post_lines}(c). During this period of time the electrons move towards the electrode at $x$ = 0 mm. The electric field changes sign again at $t \approx$ 15 ns, and acquires small negative value over most of the gap. Consequently the electron conduction current reverses direction, too, electrons drift towards the grounded electrode (anode) again. A moderate value of power density absorbed by the electrons appears at times 13 ns $\leq t \leq$ 14 ns over most of the gap (see figure~\ref{fig:post_maps}(d)). 

The post pulse plasma dynamics that we observe is rather similar to that found in \cite{Kim}, although in that work a more significant electron power absorption was observed after the termination of the excitation voltage. The quantitative differences between the findings can be attributed to the different conditions, e.g. in \cite{Kim} a low pressure discharge ($p$ = 1 Torr) was studied and pulses were switched off during a shorter time scale.  

\subsection{Double-pulse excitation}

\label{sec:double}

\begin{figure}[ht!]
\begin{center}
\footnotesize{(a)}\includegraphics[width=0.42\textwidth]{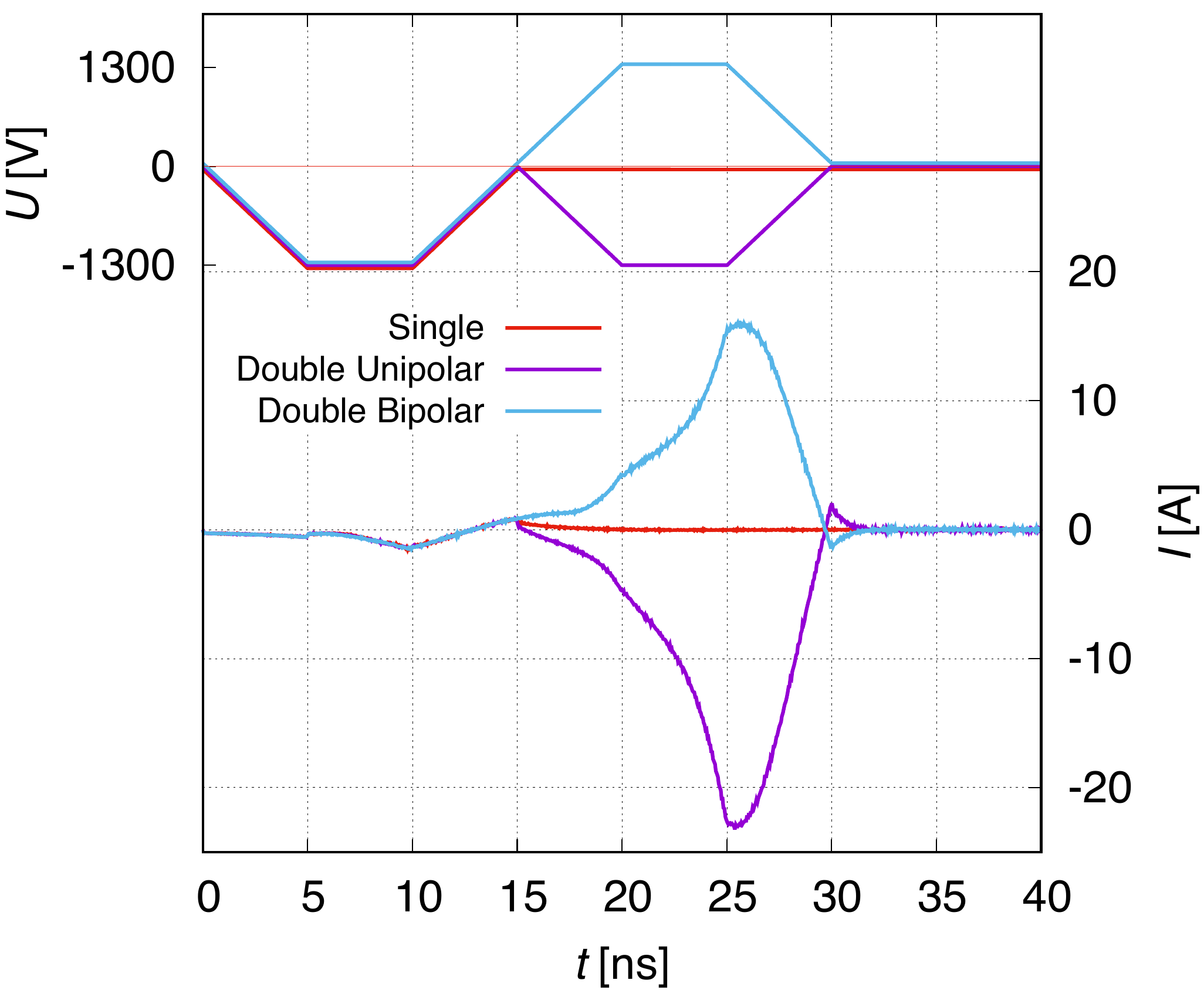}~~~~~
\footnotesize{(b)}\includegraphics[width=0.45\textwidth]{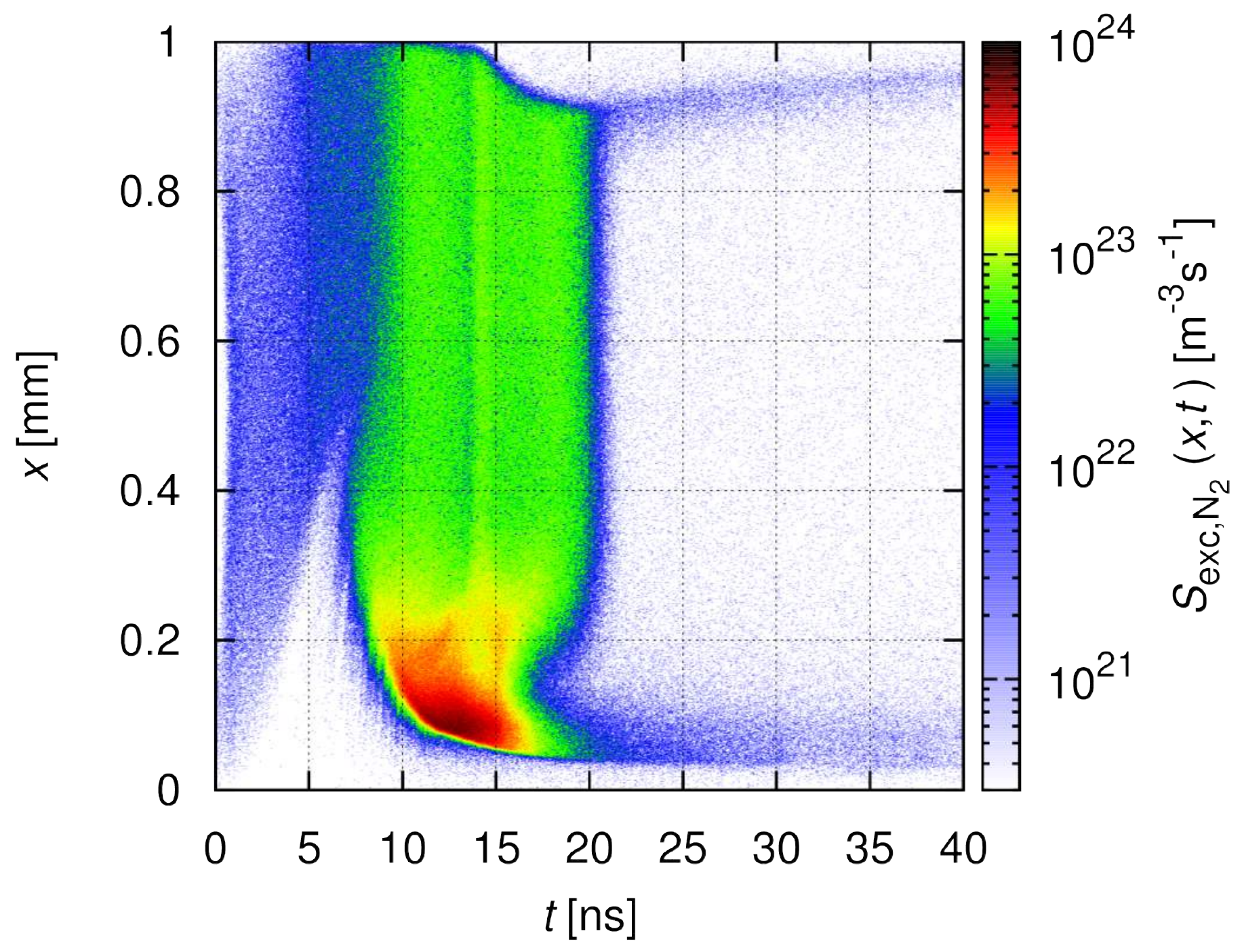}\\
\footnotesize{(c)}\includegraphics[width=0.45\textwidth]{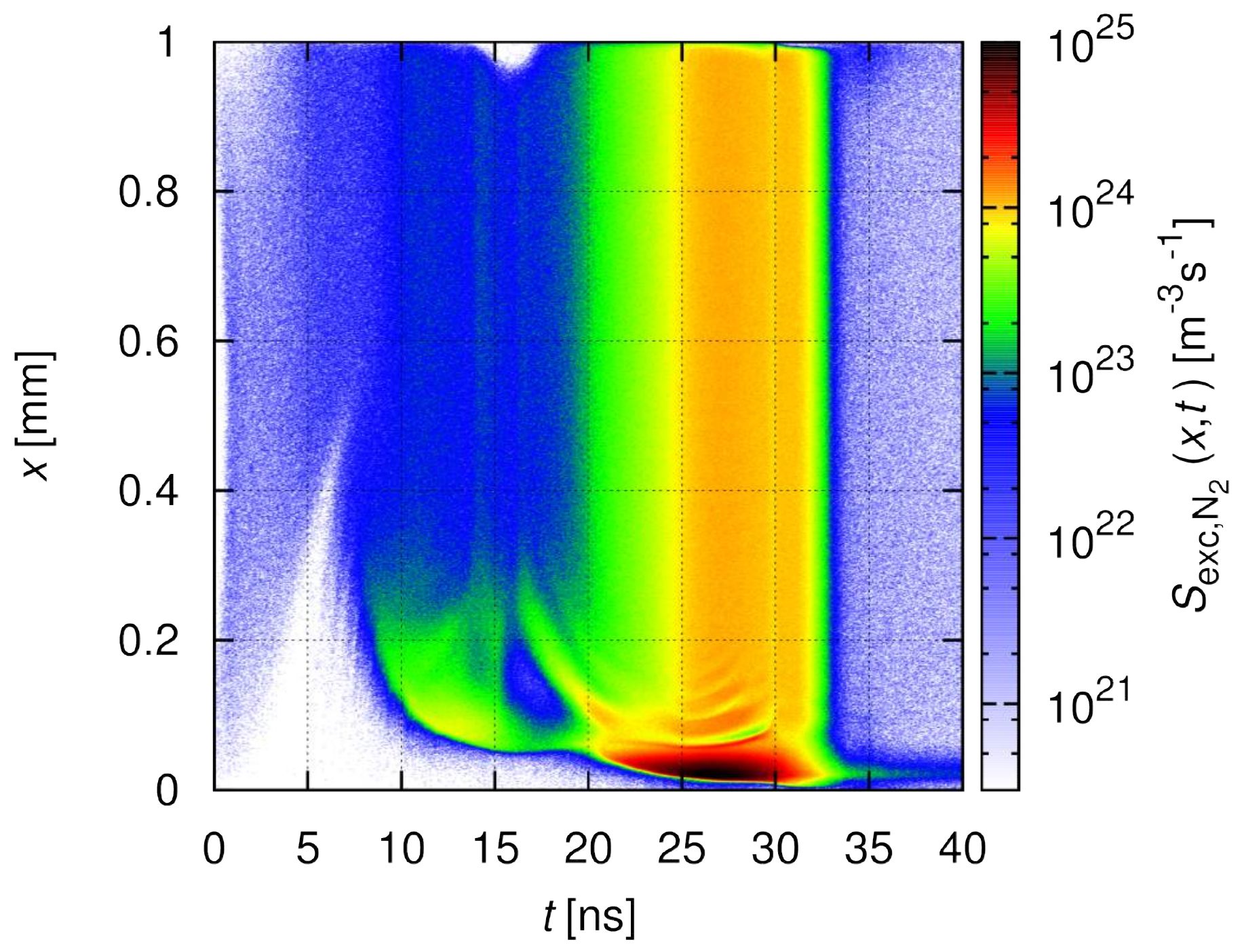}~
\footnotesize{(d)}\includegraphics[width=0.45\textwidth]{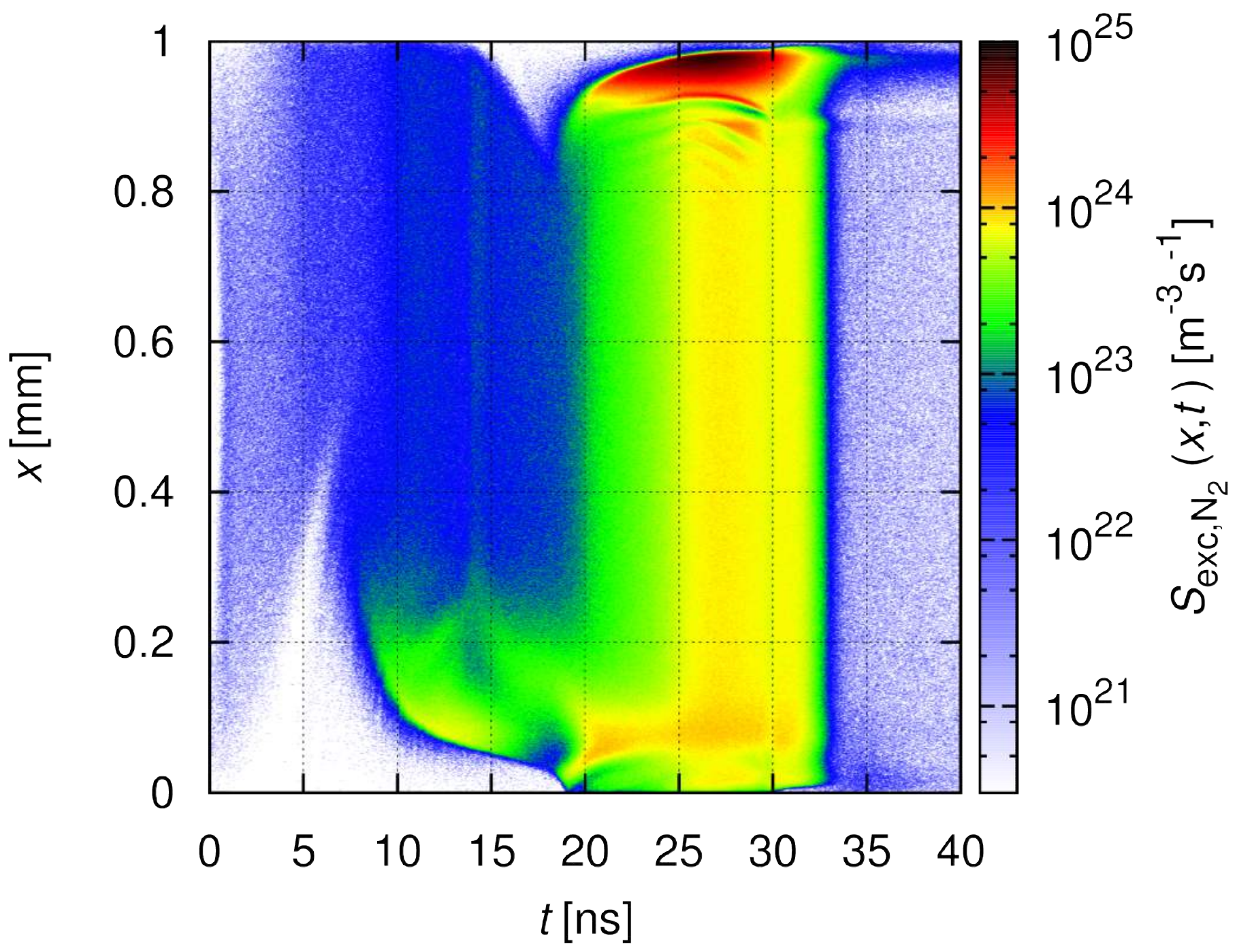}
\caption{(a) Discharge current pulses (lower set of curves, right scale) for the various excitation waveforms (upper set of curves, left scale). The spatio-temporal distribution of the excitation rate of N$_2$ molecules for (b) single-pulse, (c) unipolar double-pulse, and (d) bipolar double-pulse excitation. Discharge conditions: $U_1 = -$1300 V / $U_2 = -$1300 V, 0 V, 1300 V,  $\tau$ = 5 ns, He + 0.1\% N$_2$ at $p$ = 1 bar,  $L$ = 1 mm,  $n_0 = 1.5 \times 10^{11}$ cm$^{-3}$. (Note the different colour scales of the excitation maps.) (b--d) : the powered electrode is at $x$ = 0 mm, while the grounded electrode is at $x$ = 1 mm.)}
\label{fig:comp3}
\end{center}
\end{figure}

In this section, we first compare the behaviour of the discharges excited by single vs. double pulses. We consider the case of $U_1 = -$1300 V / $U_2 = -$1300 V, 0 V, 1300 V and $\tau$ = 5 ns (where $U_2=0$ V corresponds to the single-pulse case.). The computed current pulses for the single-pulse, as well as for unipolar and bipolar double-pulse excitation are plotted in figure~\ref{fig:comp3}. The figure also shows the spatio-temporal excitation patterns of N$_2$ molecules for the various voltage waveforms.

%source: single_uni_bip_comp/comp-currents-corr-v2.plot
%source: single_uni_bip_comp/comp-maps.plot

The single-pulse excitation produces a relatively small current at $U_0 = -$1300 V, that peaks at $I_{\rm peak} \approx -$ 1.7 A (see figure~\ref{fig:comp3}(a)), whereas the double-pulse excitation waveforms result in significantly higher currents. In the unipolar case the current peaks at a significantly higher value, $I_{\rm peak} \approx -$ 23 A. In the bipolar case the peak of the pulse has an opposite polarity and its magnitude is somewhat lower ($I_{\rm peak} \approx$ 16 A), as compared to magnitude of the peak current in the unipolar case. Panels (b), (c), and (d) of figure~\ref{fig:comp3} show the (total) excitation rate of N$_2$ molecules with spatial and temporal resolution, for the single-, unipolar double-, and bipolar double-pulse excitation, respectively. Taking, e.g., the column region of the plasma, we observe about 20 times more intensive excitation in the cases with double-pulse excitation as compared to the single-pulse case. The excitation rate in the bipolar double-pulse case is somewhat smaller, compared to the unipolar double-pulse case, because in the bipolar case the structure of the discharge needs to be "re-organised", i.e. the sheath region originally located near the electrode at $x$ = 0 mm needs to be formed at the other electrode at times $t >$ 15 ns.

\begin{figure}[ht!]
\begin{center}
\includegraphics[width=0.46\textwidth]{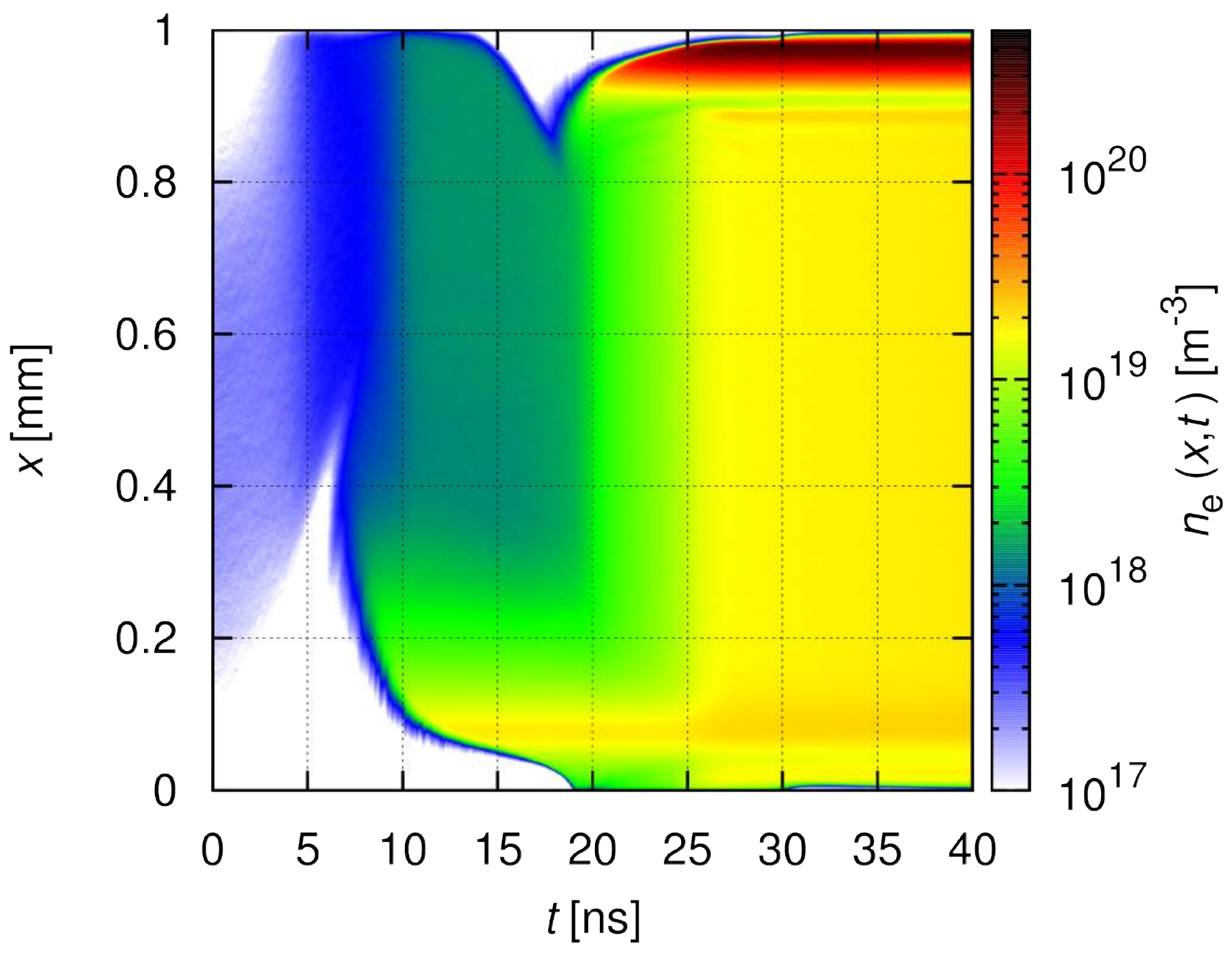}
\caption{Spatio-temporal evolution of the electron density in the case of the bipolar double-pulse excitations, at $U_0 = -$1300 V,  $\tau$ = 5 ns. (Other conditions: He + 0.1\% N$_2$ at $p$ = 1 bar,  $L$ = 1 mm,  $n_0 = 1.5 \times 10^{11}$ cm$^{-3}$.) The powered electrode is at $x$ = 0 mm, while the grounded electrode is at $x$ = 1 mm.}
\label{fig:comp3-ne}
\end{center}
\end{figure}

In both double-pulse cases the first pulse serves for creating the pre-ionisation for the second pulse. At the beginning of the simulations (as already mentioned) we set an average electron density at a value of $n_0 = 1.5 \times 10^{11}$ cm$^{-3}$. For the conditions considered here, this value grows by an order of magnitude by the end of the first pulse, as it is illustrated for the case of the bipolar pulse in figure~\ref{fig:comp3-ne}. For higher voltage amplitudes this increase can be much greater, the relatively low voltage amplitude was chosen here to avoid exceedingly high electron densities following the application of the second pulse (with the same amplitude as the first pulse).

We note that the amount of charge left in the gap by the first pulse has a very strong effect on the current amplitude during the second pulse. Similarly to this, the current created during the first pulse depends strongly on the initial charge density (assumed in the simulation or left behind from the previous pulse in a repetitively-pulsed experimental system). This effect was previously analysed in \cite{Dns}.

%source: single_uni_bip_comp/bipolar-electron-density.plot

\begin{figure}[ht!]
\begin{center}
\includegraphics[width=0.42\textwidth]{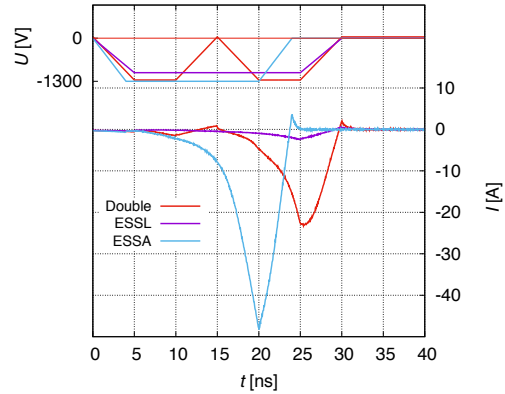}
\caption{Comparison of discharge current pulses resulting from unipolar double-pulse excitation (curves "Double", total pulse duration 30 ns, voltage amplitude $-$1300 V) and from single-pulse excitation with the same area below the $U(t)$ function as in the double-pulse case. "ESSL" denotes "Equivalent Single pulse with Same Length" (pulse duration 30 ns, voltage amplitude $-$1040 V), while "ESSA" denotes "Equivalent Single pulse with Same Amplitude" (pulse duration 24 ns, voltage amplitude $-$1300 V). (He + 0.1\% N$_2$ at $p$ = 1 bar,  $L$ = 1 mm,  $n_0 = 1.5 \times 10^{11}$ cm$^{-3}$.) Voltage: upper set of curves, left scale, current: lower set of curves, right scale.}
\label{fig:comp-shapes}
\end{center}
\end{figure}

An additional comparison between the cases of single- and double-pulse excitations is presented in figure~\ref{fig:comp-shapes}. For this comparison we take as the "base" case a double-pulse excitation with total pulse duration 30 ns ($\tau$ = 5 ns) and a voltage amplitude $U_1=U_2= -1300$\,V. The corresponding voltage and current waveforms are shown lines labelled as "Double" in figure~\ref{fig:comp-shapes}. The current reaches, for these conditions, a peak value of $\approx -$ 23 A. This type of excitation is compared to "equivalent" single-pulse excitation cases, defined as having the same $\int U(t)\,{\rm d}t$ integral. We chose two of such possible waveforms: (i) we keep the pulse duration the same (30 ns) and decrease the amplitude to $-$1040 V, resulting in an "Equivalent Single pulse with Same Length (ESSL)" and (ii) keep the amplitude at $-$1300 V and decrease the pulse length to 24 ns, resulting in an "Equivalent Single pulse with Same Amplitude (ESSA)". As figure~\ref{fig:comp-shapes} reveals, the three cases result in very different current pulse amplitudes. As compared to the single-pulse excitation, in the "ESSL" case (with decreased voltage amplitude) an about ten times lower current peak is generated. On the other hand, in spite of the fact that the duration of the excitation pulse is decreased, a factor of two higher current is generated in the ESSA case, compared to the single-pulse case. This comparison shows that the integral of the voltage waveform is by far not the principal factor in defining the peak current.

%source: modshapes-v2.plot

\begin{figure}[ht!]
\begin{center}
\footnotesize{(a)}\includegraphics[width=0.44\textwidth]{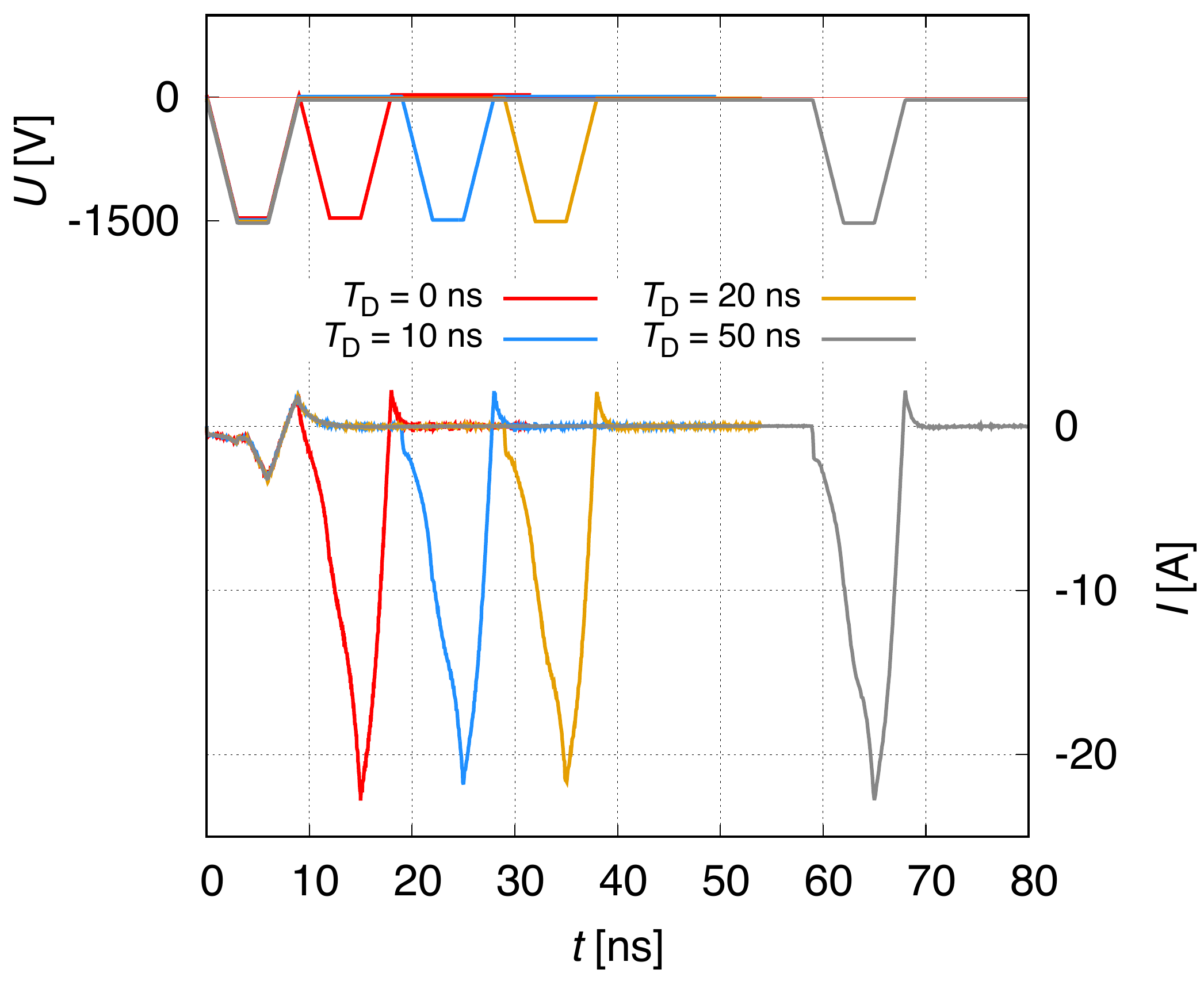}\\
\footnotesize{(b)}\includegraphics[width=0.47\textwidth]{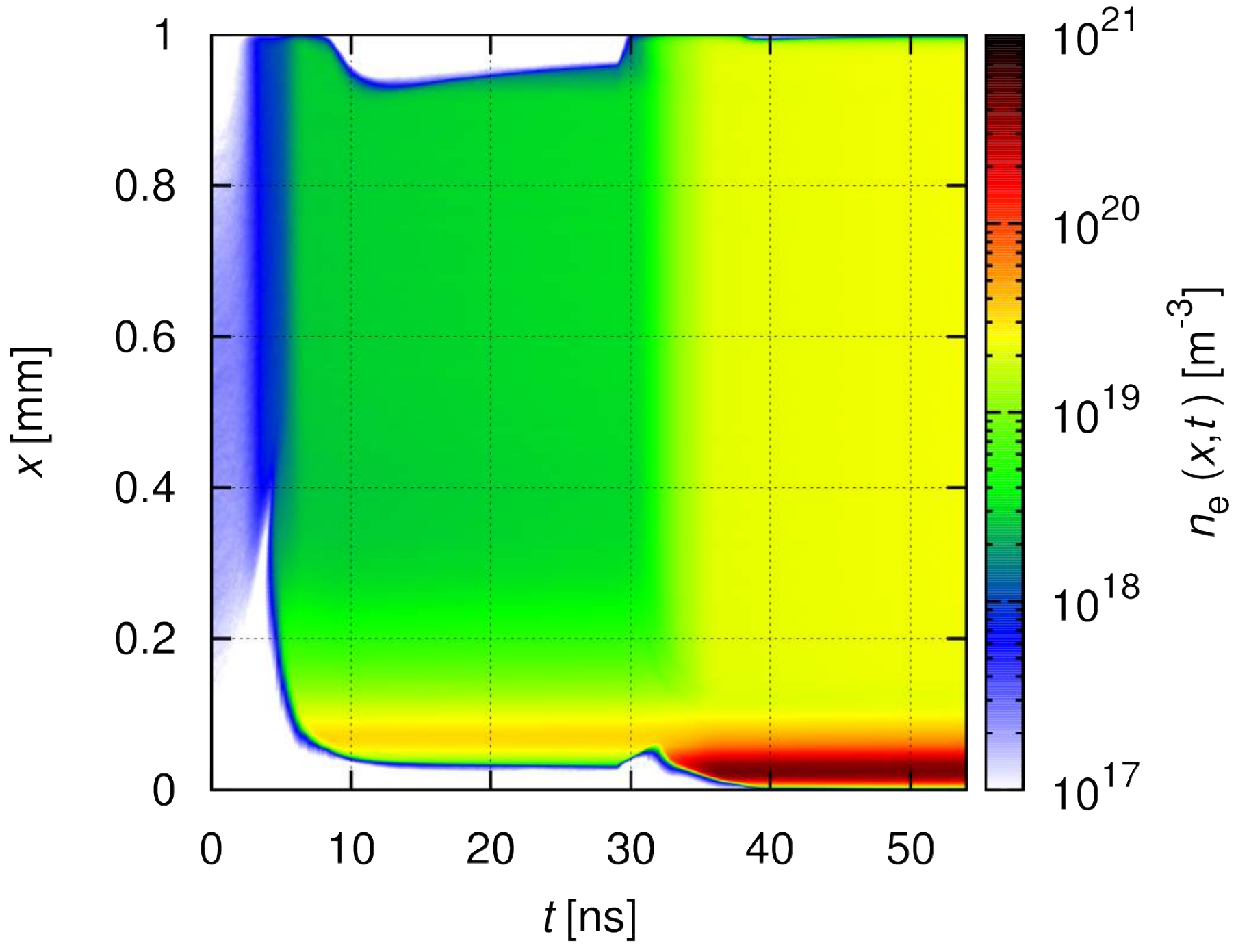}
\caption{(a) Current pulses (lower set of curves, right scale) generated by unipolar double-pulse excitation (upper set of curves, left scale) with different delay times ($T_{\rm D}$), for $U_1=U_2 = -$1500 V and $\tau$ = 3 ns. (b) Spatio-temporal distribution of the electron density for $T_{\rm D}$ = 20 ns. The powered electrode is at $x$ = 0 mm, while the grounded electrode is at $x$ = 1 mm.}
\label{fig:delay}
\end{center}
\end{figure}

Figure~\ref{fig:delay}(a) shows the current pulses obtained with double-pulse unipolar excitation, with varying delay time, $T_{\rm D}$, between the voltage pulses. For this set of simulations the peak voltage has been set at $U_1=U_2 = -$1500 V and $\tau$ was chosen to be 3 ns. The delay time is varied between 0 ns and 50 ns.

%source: effect-of-delay/comp-current2-v2.plot
%source: effect-of-delay/edensity.plot

We find that the peak current is insensitive to the delay time between the pulses. This can be explained by the fact that the decay of the plasma created by the first voltage pulse is negligible at these time scales. This is indeed found in the plot of the spatio-temporal distribution of the electron density, shown in figure~\ref{fig:delay}(b), for a delay time $T_{\rm D}$ = 20 ns. Between the two pulses, i.e. at times 9 ns $\leq t \leq$ 29 ns the $n_{\rm e}(x,t)$ distribution changes only marginally. Consequently, at delay times of the order of 10 nanoseconds the second pulse builds up a higher density plasma from the pre-ionised state created by the first pulse. The parameters of this plasma do not depend significantly on the delay time. At much longer delay times the decay of the plasma via particle fluxes to the electrodes and recombination losses may become influential, however, this is expected to occur on a significantly longer time scale.

\begin{figure}[ht!]
\begin{center}
\includegraphics[width=0.46\textwidth]{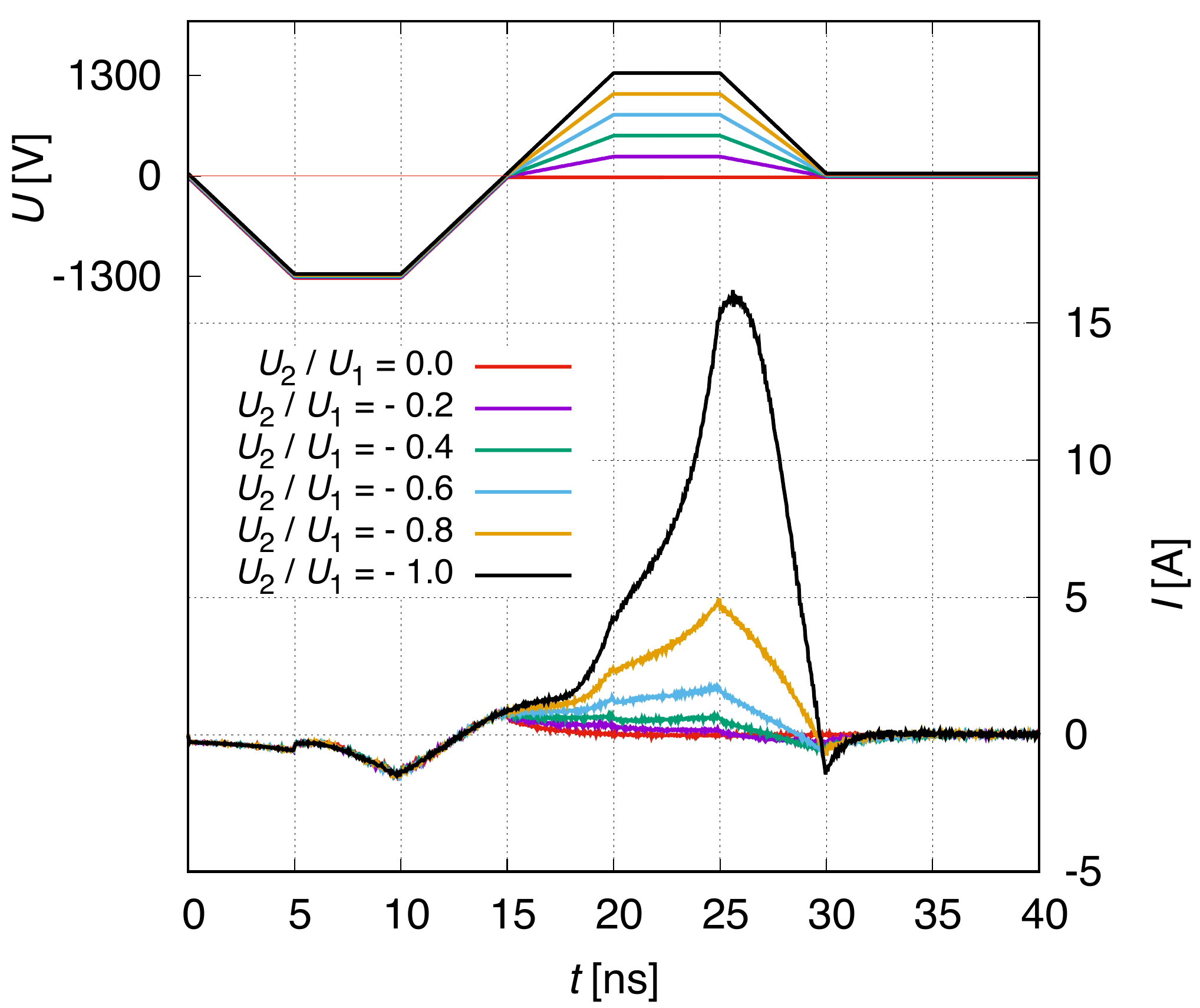}
\caption{The effect $U_2 / U_1$ ratio for bipolar double-pulse excitation at $U_1 = -$ 1300 V and $\tau$ = 5 ns. Other conditions: He + 0.1\% N$_2$ at $p$ = 1 bar,  $L$ = 1 mm,  $n_0 = 1.5 \times 10^{11}$ cm$^{-3}$. Voltage: upper set of curves, left scale, current: lower set of curves, right scale.}
\label{fig:U2}
\end{center}
\end{figure}
%source: effects/effect-of-V2-v2

Next, we address the effect of the amplitude of the second pulse, while maintaining the amplitude of the first pulse at a constant value, for bipolar double pulses. The example shown in figure~\ref{fig:U2} is for $U_1 = -1300$ V and $\tau$ = 5 ns. The findings here are very similar to the case of the effect of the voltage pulse amplitude for single pulse excitation, as the seed charged particle density in the case of single pulses and the charged particle density generated here by the first pulse play the same role. While the seed density used in both cases is $n_0 = 1.5 \times 10^{11}$ cm$^{-3}$, the peak electron density following the first pulse here is "measured" here to be about twice more than this value. Consequently, higher currents can be generated by the second pulse as compared to a single pulse with the same amplitude as the amplitude of the second pulse ($U_2$). 

\begin{figure}[ht!]
\begin{center}
\includegraphics[width=0.46\textwidth]{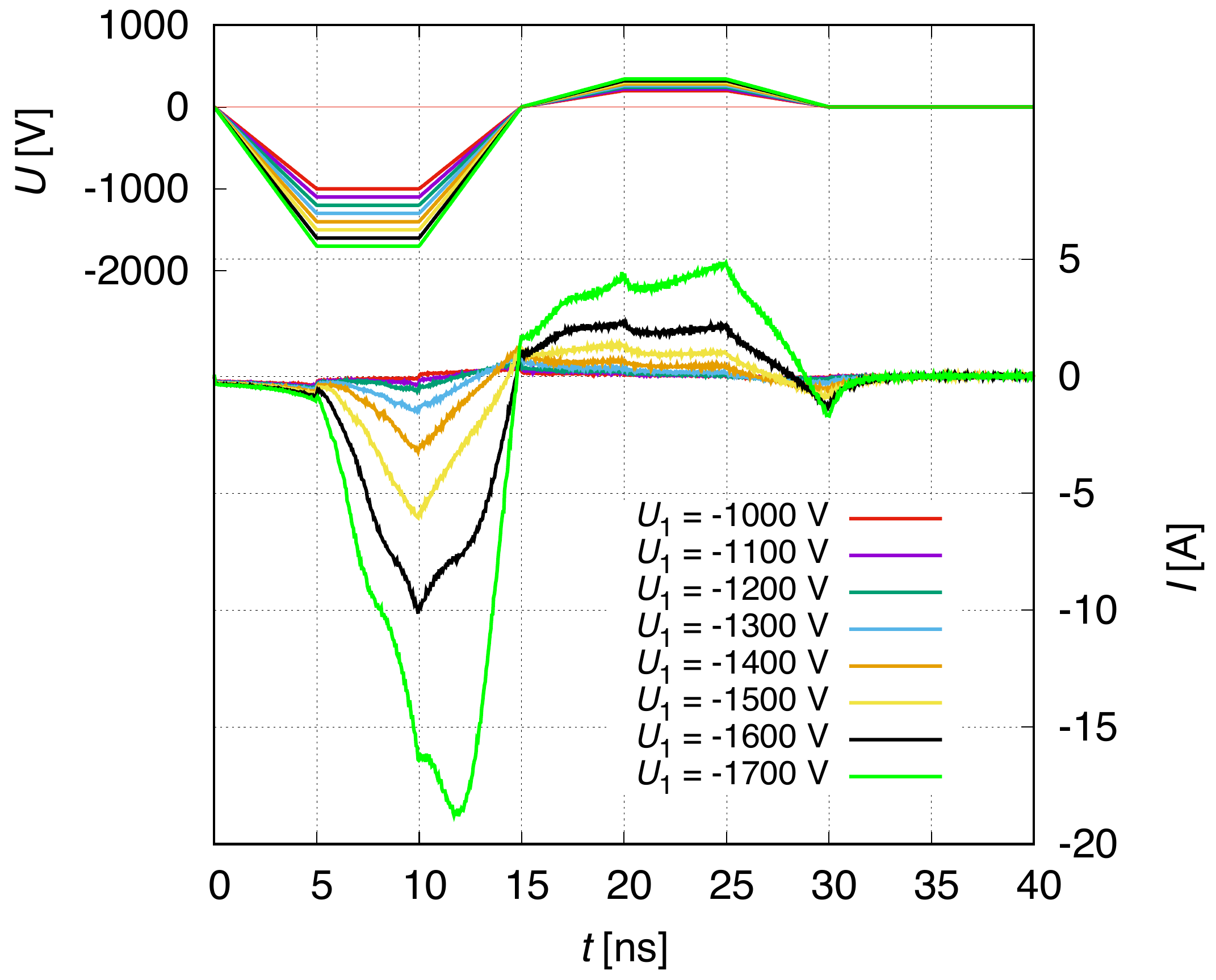}
\caption{The effect $U_1$ at fixed ratio $U_2 / U_1 = -0.2$, for $\tau$ = 5 ns, for bipolar double-pulse excitation. Other conditions: He + 0.1\% N$_2$ at $p$ = 1 bar,  $L$ = 1 mm,  $n_0 = 1.5 \times 10^{11}$ cm$^{-3}$. Voltage: upper set of curves, left scale, current: lower set of curves, right scale.}
\label{fig:U1}
\end{center}
\end{figure}
%source: effects/effect-of-V1-v2

\begin{figure}[ht!]
\begin{center}
\includegraphics[width=0.46\textwidth]{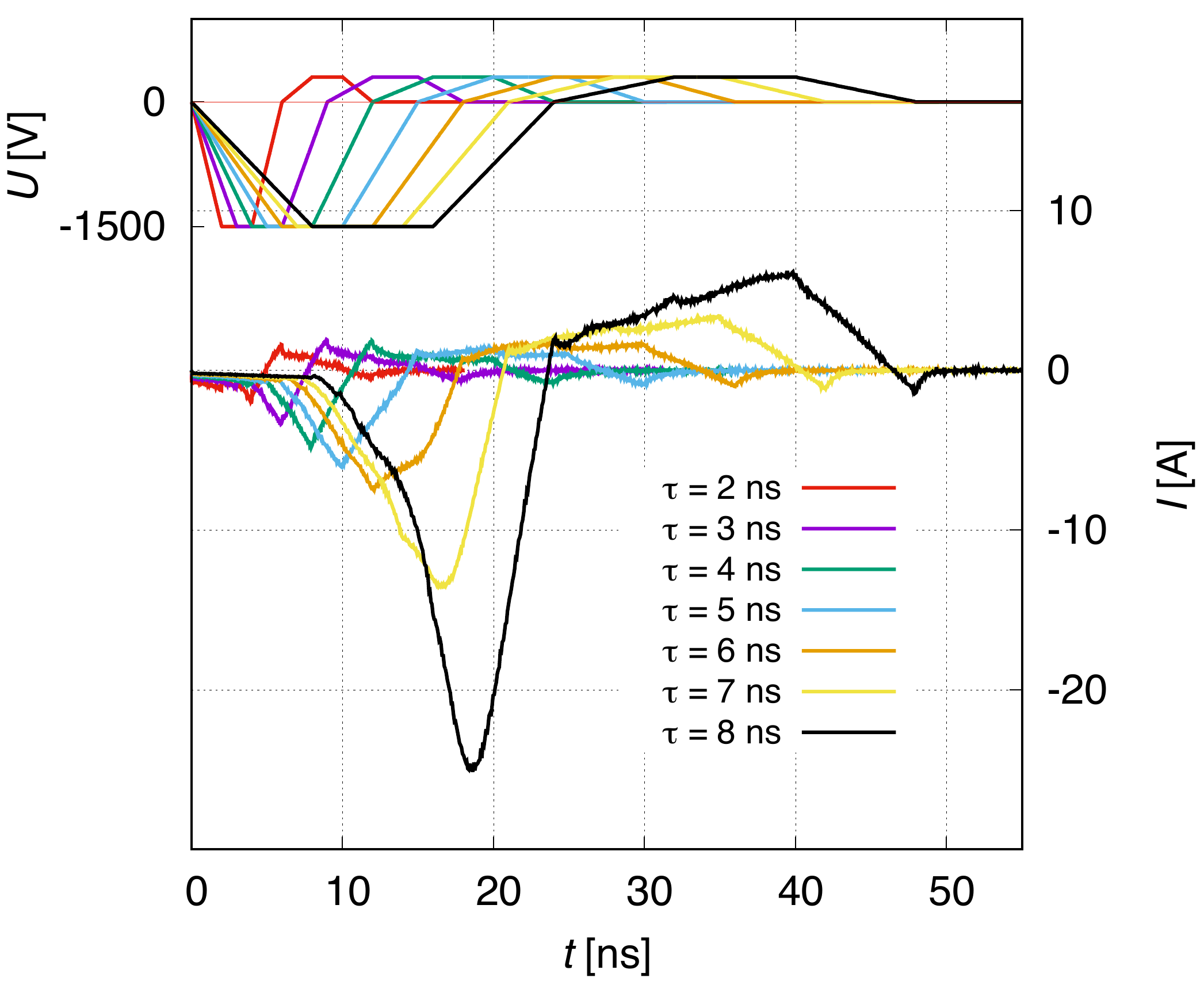}
\caption{The effect of pulse width ($\tau$) on the discharge current for $U_1 = -$ 1500 V, $U_2 = $ 300 V, for bipolar double-pulse excitation. Other conditions: He + 0.1\% N$_2$ at $p$ = 1 bar,  $L$ = 1 mm,  $n_0 = 1.5 \times 10^{11}$ cm$^{-3}$. Voltage: upper set of curves, left scale, current: lower set of curves, right scale.}
\label{fig:width}
\end{center}
\end{figure}

The choice of $U_2 / U_1=-0.2$ is investigated here, because a smaller, second voltage peak with opposite polarity with respect to the main pulse is present in some experiments, at specific conditions (e.g. \cite{Huang2015}). The effect of the voltage amplitude $U_1$ is presented in figure~\ref{fig:U1}, while the effect of the pulse length is illustrated in figure~\ref{fig:width}. Both figures reveal a strong nonlinear behaviour of the resulting current pulse shape and its peak value with the parameter varied in the respective cases.

%source: effects/effect-of-pulse-width-v2

\begin{figure}[ht!]
\begin{center}
\includegraphics[width=0.46\textwidth]{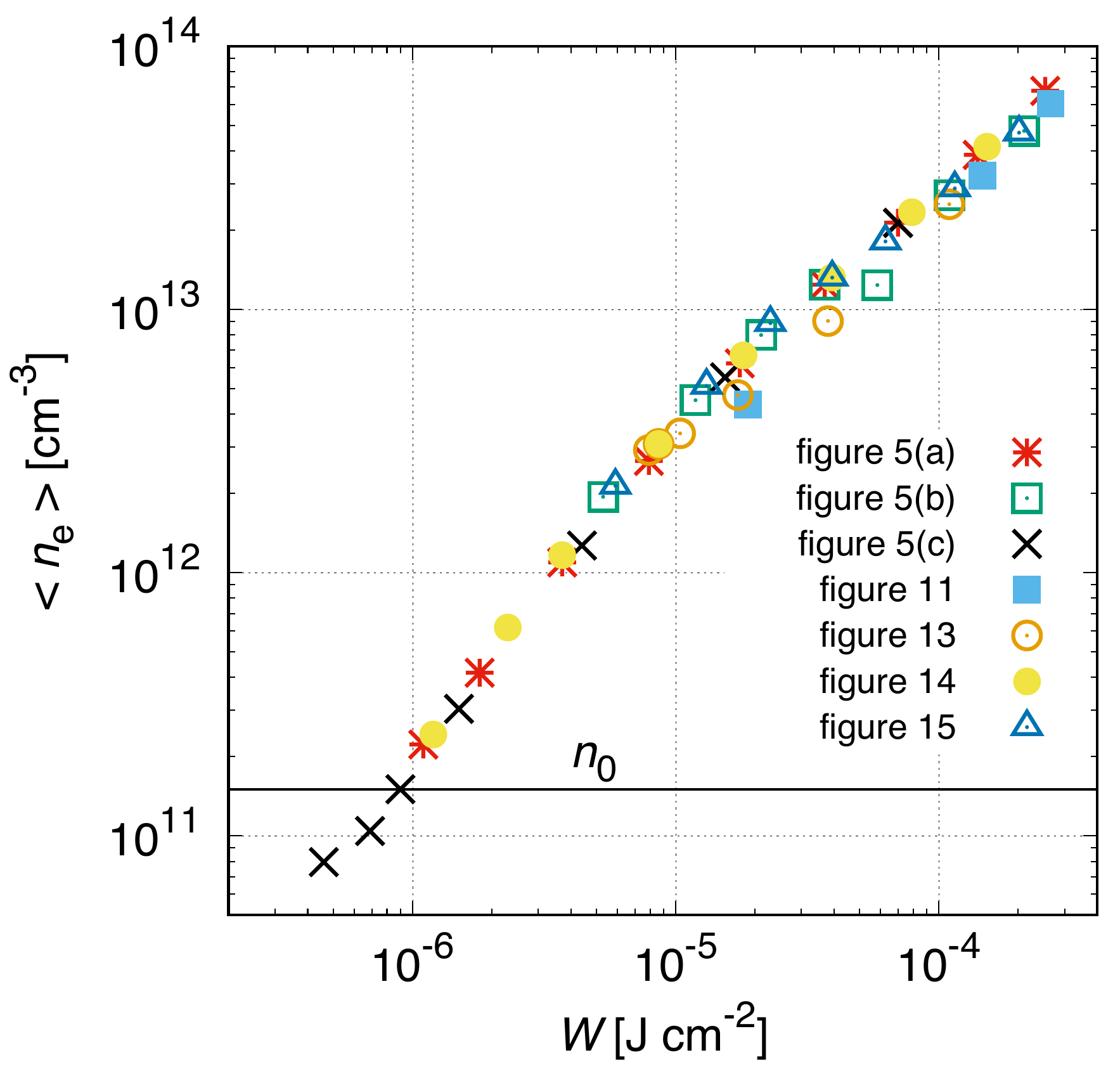}
\caption{Spatially averaged electron density at the time of the termination of the voltage pulse, as a function of the electrical energy deposited into the plasma normalised by the electrode area. The thin horizontal line marks the initial electron density $n_0 = 1.5 \times 10^{11}$ cm$^{-3}$. The symbols correspond to cases shown in previous figures, as indicated by the legend. (He + 0.1\% N$_2$ at $p$ = 1 bar,  $L$ = 1 mm).} 
\label{fig:power}
\end{center}
\end{figure}

Figure \ref{fig:power} shows the spatially averaged electron density at the time of the termination of the voltage pulse, as a function of the electrical energy deposited into the plasma, normalised by the electrode area, $W = \frac{1}{A} \int U(t) I(t) {\rm d} t$. The values are computed for one excitation pulse. The figure contains data for different waveforms and for different parameter variations, shown in previous figures. Remarkably, the data closely follow a universal curve, over a wide range of input energies, meaning that the electron density strongly depends on the electrical energy input, but it is surprisingly insensitive to the excitation voltage shape. The few data points corresponding to a final electron density that is lower than the initial density ($n_0$) originate from simulations with "stretching" a single excitation pulse: for long pulses the amplitude decreased below the breakdown voltage (as the product of the pulse length and amplitude was kept constant in this sequence of runs; see table \ref{table:cases} and figure \ref{fig:single}(c)).

%source: power/power-revised

There have been many experiments on nanosecond discharges at atmospheric pressures, with different gases and various electrode configurations. Among these, the conditions in Ref. \cite{Huang2015} are perhaps the closest to those considered in our work: in these experiments, a plane-parallel electrode configuration was used with the same gap length and helium was used as filling gas at the same pressure. The main focus of these experiments was the investigation of voltage rise time on the plasma characteristics: rise times between 0.17 kV ns$^{-1}$ and 0.42 kV ns$^{-1}$ were covered. In our studies, the voltage rise was in the same order, e.g. when investigating the effect of the pulse length (with equal rise time, plateau duration, and fall time, see figure 5(b)) the voltage rise time ranged between 0.19 kV ns$^{-1}$ and 0.75 kV ns$^{-1}$. In the experiment the discharge was ignited with 5 kHz repetition rate, while our simulations include only one pulse during which the plasma develops from a pre-defined initial charge density ($n_0 = 1.5 \times 10^{11}$ cm$^{-3}$ in this work). The value of the initial density is not known in the experiment. Nonetheless, we simulated a case shown in figure 2 of Ref [45] (with red lines). The voltage pulse was approximated with $\tau_1$ = 11 ns (rise time), $\tau_2$ = 2 ns, $\tau_3$ = 27 ns (fall time), and $U_0$ = 1800 V (peak voltage). This case corresponds to $\approx$ 0.17 kV ns$^{-1}$ voltage rise rate. The measured peak current was 5 A, corresponding to 40 A cm$^{-2}$ (according to the 0.126 cm$^2$ electrode area in the experiment). Our simulations for the same conditions give about 115 A cm$^{-2}$ current density, which is almost three times higher than in the experiment. This significant difference indicates that most likely the initial charge density is lower in the experiments than our $n_0$ value, as the plasma density and the peak current sensitively depend on the initial charge density, as it was seen, e.g., in \cite{Dns}. Despite of this discrepancy in the peak current value, we trust that our code has a predictive ability, as essentially the same code has been used for the description of radiofrequency driven atmospheric pressure microplasma jets, and an excellent agreement was found between the experimental and computed characteristics of these plasma sources \cite{Bishoff,Gibson}.

Finally we note that while throughout our studies we considered conducting electrodes, the case of dielectric electrodes can also be addressed by our simulation code. With proper modification of boundary conditions it can be used for describing the insulating surfaces with the effects resulting from surface charge accumulation. These effects are expected to have a major influence on the plasma parameters, as due to the self-terminating nature of dielectric discharges the power can only be coupled into the system during a short time. Therefore it is expected that, e.g., higher voltages will be needed to establish the same plasma density as obtained with conducting electrodes. Such studies are planned as future work.

\section{Summary}

In this work, we have investigated the effect of the pulse shape on the characteristics of nanosecond discharges created in He + 0.1\% N$_2$ mixture, at atmospheric pressure. We have studied the cases of single-pulse and double-pulse excitation. 

In the case of single-pulse excitation we addressed the effects of voltage amplitude and pulse duration, and analysed in detail the dynamics of the plasma following the termination of the excitation voltage pulse. The latter was found to be governed by the effect of space charges that generate an electric field reversal past the excitation pulse, and result in the development of a reverse current peak and later on in the establishment of ambipolar electric fields in the vicinity of the electrodes. These transient effects were found to be largely controlled by the fall-time of the excitation pulse. 

In the case of double-pulse excitation we investigated (i) the differences of the plasma behaviour between unipolar and bipolar excitation pulses, (ii) the behaviour of discharges created by double vs. "equivalent" single voltage pulses, (ii) the effect of delay time between the two pulses (in the unipolar case), (iii) the effect of voltage amplitude(s) and pulse length in the case of bipolar excitation pulses that approximate experimentally relevant waveforms.

Analysing all the investigated cases together, we have found a very strong correlation between the electrical input energy into the plasma and the electron density (measured at the end of the excitation pulse). This analysis has also pointed out that the excitation voltage waveform, on the other hand, has a weak effect on the electron density, as long as the input energy is kept to be the same.

\ack This work was supported by the National Office for Research, Development and Innovation (NKFIH) via grant 119357, Osaka University International Joint Research Promotion Program (Type A), the JSPS Grants-in-Aid for Scientific Research (S) 15H05736, and the UK EPSRC (EP/K018388/1) grant.

\end{document}